\documentclass[journal=jctcce,manuscript=article, layout=twocolumn]{achemso}

\usepackage{amsmath, amssymb}
\usepackage[utf8]{inputenc}
\usepackage{graphicx}
\usepackage{color}
\usepackage{comment}
\usepackage{bm}
\usepackage[breaklinks=true]{hyperref}
\usepackage{physics}
\usepackage{braket}
\usepackage[version=4]{mhchem}
\usepackage{enumitem}   
\usepackage{subfig}
\usepackage{mathtools}
\usepackage{cuted}

\newcommand{\mr}[1]{\mathrm{#1}}
\newcommand{\mc}[1]{\mathcal{#1}}
\newcommand{\bmth}{{\bm{\theta}}}
\newcommand{\bmkp}{{\bm{\kappa}}}

\date{\today}

\title{Analytical energy gradient for state-averaged orbital-optimized variational quantum eigensolvers and its application to a photochemical reaction}
\author{Keita Omiya}
\email{keita.arimitsu@psi.ch}
\affiliation{QunaSys Inc., Aqua Hakusan Building 9F, 1-13-7 Hakusan, Bunkyo, Tokyo 113-0001, Japan}
\alsoaffiliation{Department of Physics, ETH Z\"{u}rich, CH 8093 Z\"{u}rich, Switzerland}
\alsoaffiliation{Condensed Matter Theory Group, LSM. NES, Paul Scherrer Institute, Villigen PSI CH-5232, Switzerland}
\alsoaffiliation{Institute of Physics, Ecole Polytechnique F\'ed\'erale de Lausanne (EPFL), CH-1015 Lausanne, Switzerland}
\author{Yuya O. Nakagawa}
\email{nakagawa@qunasys.com}
\affiliation{QunaSys Inc., Aqua Hakusan Building 9F, 1-13-7 Hakusan, Bunkyo, Tokyo 113-0001, Japan}

\author{Sho Koh}
\affiliation{QunaSys Inc., Aqua Hakusan Building 9F, 1-13-7 Hakusan, Bunkyo, Tokyo 113-0001, Japan}

\author{Wataru Mizukami}
\affiliation{Center for Quantum Information and Quantum Biology,
Institute for Open and Transdisciplinary Research Initiatives, Osaka University, Japan.}
\alsoaffiliation{JST, PRESTO, 4-1-8 Honcho, Kawaguchi, Saitama 332-0012, Japan}
\alsoaffiliation{Graduate School of Engineering Science, Osaka University, 1-3 Machikaneyama, Toyonaka, Osaka 560-8531, Japan.}

\author{Qi Gao}
\affiliation{Mitsubishi Chemical Corporation, Science \& Innovation Center, 1000, Kamoshida-cho, Aoba-ku, Yokohama 227-8502, Japan}

\author{Takao Kobayashi}
\affiliation{Mitsubishi Chemical Corporation, Science \& Innovation Center, 1000, Kamoshida-cho, Aoba-ku, Yokohama 227-8502, Japan}

\begin{document}

\begin{abstract}
Elucidating photochemical reactions is vital to understand various biochemical phenomena and develop functional materials such as artificial photosynthesis and organic solar cells, albeit its notorious difficulty by both experiments and theories.
The best theoretical way so far to analyze photochemical reactions at the level of ab initio electronic structure is the state-averaged  multi-configurational self-consistent field (SA-MCSCF) method. 
However, the exponential computational cost of classical computers with the increasing number of molecular orbitals hinders applications of SA-MCSCF for large systems we are interested in.
Utilizing quantum computers was recently proposed as a promising approach to overcome such computational cost, dubbed as state-averaged orbital-optimized variational quantum eigensolver (SA-OO-VQE).
Here we extend a theory of SA-OO-VQE so that analytical gradients of energy can be evaluated by standard techniques that are feasible with near-term quantum computers.
The analytical gradients, known only for the state-specific OO-VQE in previous studies, allow us to determine various characteristics of photochemical reactions such as the conical intersection (CI) points.
We perform a proof-of-principle calculation of our methods by applying it to the photochemical {\it cis-trans} isomerization of 1,3,3,3-tetrafluoropropene.
Numerical simulations of quantum circuits and measurements can correctly capture the photochemical reaction pathway of this model system, including the CI points.
Our results illustrate the possibility of leveraging quantum computers for studying photochemical reactions.
\end{abstract}

\maketitle
\section{\label{sec:introduction}Introduction}
Vision~\cite{Fernald2006Science, Schnedermann2018}, light tolerance of DNAs and proteins~\cite{Ismail2002, Middleton2009ARPC}, light-harvesting in photosynthesis~\cite{Cheng2009ARPC}, and luminescence of fireflies~\cite{Fraga2008firefly}: these are examples of biochemical phenomena induced by light. Industrially important materials based on photophysical and photochemical processes are, say, OLEDs~\cite{Song2020AdMat}, artificial photosynthesis~\cite{Karkas2014ChemRev}, photocatalysts~\cite{Melchionna2020acscatal}, solar cells~\cite{McEvoy2012Book}, fluorescent probes~\cite{Ueno2011NatureMethods}, and photoresists~\cite{Li2017CSR}. Although those are not exhaustive lists, they illustrate the importance of understanding photophysical and photochemical processes. A detailed understanding of these processes, especially for photochemical reactions, is essential to answer how living things work at molecular levels and rationally design materials for achieving a sustainable society, such as highly-efficient solar cells and artificial photosynthesis. 

Despite its importance, the analysis of photochemical reactions at the atomic and electronic levels still remains a challenge for both experiment and theory. This difficulty stems from the fact that several energetically close quantum states could be involved in a photochemical reaction. Indeed, it is relatively easy to obtain information in the Franck-Condon region, where energy gaps are large and light-absorptions (or emissions) occur. On the other hand, nonradiative processes are challenging to observe because the energy gap is tiny or zero at a critical point, as in the case of conical intersections (CIs). Throughout this paper, we use the term ``CI" to mean the conical intersection, not the configuration interaction. Photochemical processes via CIs are extremely fast and require an ultrashort-wave light source for experimental observations. In recent years, x-ray free-electron lasers have made it possible to obtain ultrashort pulses~\cite{Maiuri2019JACS}. Thanks to them, it is now getting possible to observe outcomes of the existence of CIs with spectroscopic techniques for small molecules~\cite{Timmers2019NatComm,Kobayashi2019Science,Zinchenko2021Science}, although CI itself cannot be measured as it is mathematical object. Nevertheless, as with most spectroscopic methods, these techniques do not provide direct and detailed information at the atomic or electronic level.

As a complement to experiments, computer simulations have played a major role in studying photochemical processes because they can provide experimentally inaccessible information, including the detailed geometry of a molecule and its changes during a reaction. However, describing the region where the two quantum states are close to each other, such as CIs, is rather difficult on computers as it is with experiments.
It is known that typical quantum chemical theories based on a single reference state, such as density functional theory (DFT) and coupled-cluster theory (CC), are not suitable for locating CIs~\cite{Benjamin2006MolPhys,Kohn2007JCP}.
Moreover, excited states may have strongly correlated electrons that require taking static correlations into account by referencing to multi-configurational self-consistent field (MCSCF)~\cite{Bernardi1990}.
A remedy for these two issues is the combination of the following two methods: one is a state-average method (SA), which optimizes several quantum states simultaneously~\cite{Docken1972lih,Hinze1973mc,Slater1960quantum,McWeeny1974SASCF}.
The SA technique allows us accurate and smooth description of potential energy surfaces (PESs) around CIs, which is difficult to obtain with the state-specific (SS) optimization.
The other is a multi-configurational (MC) method~\cite{Roos2016multiconfigurational}, which can handle strong electron correlations.  Note that MC accurately treats a limited Hilbert space, called ``active space," consisting of orbitals and electrons preselected by the user.  The most widely-used SA-MC methods are state-averaged complete active space self-consistent field (SA-CASSCF) and second-order perturbative corrections for it~\cite{Nakano1993quasidegenerate,Bernardi1996potential,Finley1998multi,Angeli2004quasidegenerate,Shiozaki2011xmscaspt2,Granovsky2011extended}.

The primary problem of (SA-)MC is that its computational costs severely limit the active space's size; SA-CASSCF's computational costs increase exponentially with the number of electrons and orbitals of the active space.
Today's SA-CASSCF calculations by classical computers can only handle a few dozen active spaces at most. 
Such a small active space does not take into account the electron correlations necessary for quantitative discussions, which are often called dynamical correlations.
Practically, at this moment, one has no choice but to use a second-order perturbation theory such as CASPT2 or NEVPT2 to consider the dynamical electron correlation.
Various alternative methods such as RASSCF~\cite{Malmqvist1990restricted}, DMRG-CASSCF\cite{Ghosh2008orbital}, Full-CI QMC~\cite{Booth2009fermion}, and heat-bath CI~\cite{Holmes2016heat, Sharma2017semistochastic}  have been proposed to alleviate this limitation.
Still, they have not yet been able to replace SA-CASSCF.


Here, we present a new route to tackle photochemical reactions using a quantum computer, especially near-term quantum computer which is called Noisy-Intermediate-Scale Quantum (NISQ) device~\cite{Preskill2018}.
NISQ devices can manipulate simple circuits involving typically several hundred of quantum bits (qubits) without error-correction.
Although NISQ devices cannot execute complicated quantum circuits (computations), they still has a potential of outperforming any existing classical computers~\cite{Arute2019,Zhong1460}.
Our method is based on a hybrid quantum-classical algorithm known as the variational quantum eigensolver (VQE)~\cite{peruzzo2014variational}. The VQE uses both quantum and classical computers and efficiently handles the superposition of electrons in the active space. Several groups, including ours, have already developed a ``CASSCF"  (to be more precise, MCSCF)  method based on the VQE on quantum computers, called the orbital-optimized VQE (OO-VQE)~\cite{Takeshita2020, Mizukami2020,Sokolov2020JCP}.
Afterwards, the state-averaged version of OO-VQE (SA-OO-VQE) was also  proposed~\cite{Yalouz_2021}.
However, its analytical energy derivatives has not been available yet, despite its important role in analyzing photochemical reactions by determining minimum energies, transition states, CI points, reaction paths, etc. In this study, we have formulated and implemented analytical energy derivatives of SA-OO-VQE.
As a proof-of-principle, the algorithm is applied to the photochemical {\it cis-trans} isomerization of 1,3,3,3-tetrafluoropropene (TFP). We have successfully computed the photochemical reaction pathway of this model system, including the minimum energy CI.
Our results open up a way towards quantum computational analyses of photochemical reactions.
\section{Setup \label{sec:preliminaries}}
We consider the Hamiltonian for electronic states of a given molecule that depends on parameters $\bm{x}$ such as nuclear coordinates of the molecule or static electromagnetic field.
The second-quantized form of the Hamiltonian, which is suitable for quantum computers to deal with~\cite{mcardle2018quantum, Cao2018}, is
\begin{equation}
\label{eq: full-space hamiltonian}
\begin{split}
 \hat{H}(\bm{x}) &= E_\mr{c}(\bm{x}) + \sum_{ij,\sigma} h_{ij}(\bm{x})\hat{a}_{i\sigma}^\dag\hat{a}_{j\sigma}
 \\ &+ \frac{1}{2}\sum_{ijkl,\sigma\tau}g_{ijkl}(\bm{x})\hat{a}^\dag_{i\sigma}\hat{a}^\dag_{j\tau}\hat{a}_{k\tau}\hat{a}_{l\sigma},
\end{split}
\end{equation}
where $E_\mr{c}(\bm{x})$ is a scalar depending on $\bm{x}$ (a constant contribution of the energy), $h_{ij}(\bm{x}) \: (g_{ijkl}(\bm{x}))$ is one-electron (two-electron) integral, and  $\hat{a}_{i\sigma}\left(\hat{a}^\dag_{i\sigma}\right)$ is an annihilation (creation) operator corresponding to $i$th molecular orbital (MO) with spin $\sigma$.
Those operators satisfy the fermionic anti-commutation relation $\left\{\hat{a}_{i\sigma},\hat{a}^\dag_{j\sigma'}\right\}=\delta_{ij}\delta_{\sigma\sigma'}$,
where $\left\{A, B\right\}\equiv AB+BA$ is the anti-commutator and $\delta$ is the Kronecker delta.

The wavefunction under the active space approximation is written as $\ket{\mr{vac}}_\mr{vir} \otimes \ket{\psi} \otimes\ket{\uparrow\downarrow}_\mr{core}$, where $\ket{\mr{vac}}_\mr{vir}$ is the vacuum state for virtual orbitals, $\ket{\psi}$ is a wavefunction in the active space, and $\ket{\uparrow\downarrow}_\mr{core}$ is the fully-occupied state for core orbitals.
The original Hamiltonian $H$ is projected onto the effective Hamiltonian $\hat{\mc{H}}$ in the active space defined by requiring 
$ \hat{H} \left( \ket{\mr{vac}}_\mr{vir} \otimes \ket{\psi} \otimes\ket{\uparrow\downarrow}_\mr{core} \right) = \hat{\mc{H}}\ket{\psi} $
for an arbitrary state $\ket{\psi}$ in the active space.

In (SA-)MC calculation, the MOs are optimized by tuning orbital parameters $\kappa_{pq}$ which define the rotation operator
\begin{equation} \label{eq: def OO rotation}
 \hat{U}_{\mr{OO}}(\bm{\kappa})=\exp\left[\sum_{p>q}\sum_\sigma\kappa_{pq}\left(\hat{a}_{p\sigma}^\dag\hat{a}_{q\sigma} - \hat{a}_{q\sigma}^\dag\hat{a}_{p\sigma}\right)\right].
\end{equation}
The operator $\hat{U}_{\mr{OO}}(\bm{\kappa})$ alters the original Hamiltonian $\hat{H}(\bm{x})$ to $\hat{U}_{\mr{OO}}^\dag(\bm{\kappa})\hat{H}(\bm{x})\hat{U}_{\mr{OO}}(\bm{\kappa})$.
We define the active space Hamiltonian 
$\hat{\mc{H}}(\bm{x}, \bm{\kappa})$ depending on $\bm{x}, \bm{\kappa}$ as the active space projection of $\hat{U}_{\mr{OO}}^\dag(\bm{\kappa})\hat{H}(\bm{x})\hat{U}_{\mr{OO}}(\bm{\kappa})$. 
We call the optimization of $\bm{\kappa}$ as orbital optimization (OO).

As mentioned in Introduction, 
utilizing quantum computers in (SA-)MCSCF method was proposed 
in Refs.\citenum{Mizukami2020,Takeshita2020, Sokolov2020JCP, Yalouz_2021}.
In those proposals, quantum computers play a role of an eigensolver of $\hat{\mc{H}}(\bm{x},\bm{\kappa})$, which costs exponential computational resources when using classical computers.
With a method called variational quantum eigensolver (VQE)~\cite{peruzzo2014variational} and its extensions, one can calculate (approximate) eigenvalues and eigenstates of $\hat{\mc{H}}(\bm{x},\bm{\kappa})$ by quantum computers.
The one- and two- particle reduced density matrices (1,2-RDMs), $\rho^{(1)}_{ij} = \ev{\sum_\sigma \hat{a}_{i\sigma}^\dag\hat{a}_{j\sigma}}$, $\rho^{(2)}_{ijkl} = \ev{\sum_{\sigma\tau} \hat{a}_{i\sigma}^\dag \hat{a}_{j\tau}^\dag \hat{a}_{k\tau} \hat{a}_{l\sigma}}$ , are also evaluated by quantum computers and the MOs are optimized by classical computers based on the evaluated RDMs.
In (SA-)MCSCF, the optimization of MOs  and the call of eigensolver for $\hat{\mc{H}}(\bm{x}, \bm{\kappa})$ is repeated until convergence.
We name the combination of VQE and (SA-)MCSCF as (SA-)OO-VQE.
A brief review of quantum computing and quantum chemistry is found in Supporting Information (SI).

There are various methods proposed for computing multiple eingenvalues of $\hat{\mc{H}}(\bm{x}, \bm{\kappa})$ by near-term quantum computers, which do not afford to execute complicated quantum circuits.
In this work we focus on three of them: subspace-search VQE (SSVQE)~\cite{nakanishi2018subspace}, multistate-contracted VQE (MCVQE)~\cite{Parrish2019PRL}, and variational quantum deflation (VQD)~\cite{Higgott2019vqd}.
We specifically name the SA-MCSCF calculation accompanied with SSVQE, MCVQE, and VQD as SA-OO-SSVQE, SA-OO-MCVQE, and SA-OO-VQD, respectively.
Note that SA-OO-SSVQE is called SA-OO-VQE in Ref.~\citenum{Yalouz_2021}. 
Although these SA-OO-VQEs can give accurate energy of a target system, the lack of analytic energy derivatives has hindered their applications to interesting phenomena in photochemistry. Note that the Hellman-Feynman theorem to calculate the derivative cannot be applied to eigenenergies obtained by SA-OO-VQEs because each energy is not variationally minimized with respect to $\bm{\kappa}$ (see Results section).

In the following, we explain our theory to calculate energy derivatives of SA-OO-VQE by taking SA-OO-VQD as an example because VQD may exhibit the best performance among the three~\cite{ibe2020calculating}.
The other two cases are described in SI.
\section{\label{sec:results}Results}
We first review an algorithm of SA-OO-VQD.
VQD is an iterative method to obtain excited states of a given Hamiltonian $\hat{\mc{H}}$.
The trial state called {\it ansatz} state is set as $\ket{\psi(\bm{\theta})} = \hat{U}(\bm{\theta})\ket{\psi}$, where $\ket{\psi}$ is an initial state prepared on quantum computers and $\hat{U}(\bm{\theta})$ is a quantum circuit with parameters $\bm{\theta}$ called {\it circuit parameters}.
Let us assume that one obtains the eigenstates of $\hat{\mc{H}}$ up to $S-1$th level as $\ket{\psi(\bm{\theta}^*_0)}, \ldots, \ket{\psi(\bm{\theta}^*_{S-1})}$ with the optimized parameters $\bmth_0^*,\ldots, \bmth_{S-1}^*$.
The $S$th eigenstate can be found by minimizing the cost function 
\begin{equation}\label{eq:vqd}
\begin{split}
    F_{S}^{\mr{VQD}}(\bm{\theta}_{S})&= \bra{\psi}\hat{U}^\dag(\bm{\theta}_{S})\hat{\mc{H}}\hat{U}(\bm{\theta}_{S})\ket{\psi} \\
    &+ \sum_{T<S}\beta_T\left|\bra{\psi}\hat{U}^\dag(\bm{\theta}_{S})\hat{U}(\bm{\theta}_T^*)\ket{\psi}\right|^2
\end{split}
\end{equation}
with respect to $\bmth_S$, where $\beta_T$ is some positive number to assure the orthogonality of $\ket{\psi(\bm{\theta}_{S})}$ to all of the lower eigenstates~\cite{Higgott2019vqd}.
The quantum computer is run to evaluate Eq.~(\ref{eq:vqd}) at various parameters $\bmth_S$ and the parameters are optimized based on the evaluated values.
The optimal circuit parameters $\bmth_S^*$ represents the $S$th eigenstate by $\ket{\psi(\bm{\theta}_{S}^*)}$. 
Starting from $S=0$, we can iteratively optimize $F_{S}$ to reach the desired eigenlevel. 

In SA-OO-VQD, the MOs (i.e., the orbital parameters $\bmkp$) are optimized to minimize the state-averaged energy, $E^{\mr{SA}} = \sum_{S=0}^{K-1} \omega_S^{\mr{SA}} E_S$, where $\{E_S\}_{S=0}^{K-1}$ are $K$ lowest eigenvalues of the active space Hamiltonian $\hat{\mc{H}}(\bm{x},\bmkp)$ and $\omega_S^{\mr{SA}} > 0$ is a weight parameter for SA satisfying $\sum_S \omega_S^{\mr{SA}} = 1$.
In practice, the minimization of $E^{\mr{SA}}$ proceeds as follows:
First, with fixed $\bmkp$, the eigenstates of $\hat{\mc{H}}(\bm{x}, \bm{\kappa})$ are calculated by VQD.
The circuit parameters $\bmth_0, \ldots, \bmth_{K-1}$ are optimized to approximate the eigenstates as $\ket{E_S} \approx \ket{\psi(\bmth_S^*)}$.
Next, the orbital parameters $\bmkp$ are updated by using the 1,2-RDMs for the obtained eigenstates.
The RDMs are evaluated by quantum computers, but the update of $\bmkp$ can be done by classical computers.
These two steps are repeated until $E^{\mr{SA}}$ converges.
After the convergence, one can obtain precise eigenstates and eigenenergies of the target system with the optimized parameters $\bmkp^*$ and $\bmth_0^*, \ldots, \bmth_{K-1}^*$.
Note that $\bmkp^*$ and $\bmth_0^*, \ldots, \bmth_{K-1}^*$ depend on $\bm{x}$ as $\bmkp^*(\bm{x})$ and $\bmth_0^*(\bm{x}), \ldots, \bmth_{K-1}^*(\bm{x})$.
Throughout this paper, we consider analytical derivatives of eigenenergies of $\hat{\mc{H}}(\bm{x}, \bm{\kappa}^*(\bm{x}))$ with respect to the parameters $\bm{x}$.

We then explain our main theoretical result.
Our procedures to obtain the gradient of the energy eigenvalue of $\hat{\mc{H}}(\bm{x}, \bm{\kappa}^*(\bm{x}))$ are based on the Lagrangian method developed in quantum chemistry with classical computer\cite{Helgaker1989}.
We consider analytical gradients of energies obtained by SA-OO-VQD with respect to $x_\mu$ ($\mu$th component of $\bm{x}$).
We define the Lagrangian for $A$th state as
\begin{equation}
\begin{split}
 & L_A^\mr{VQD}(\bm{x}, \bm{\theta}_0, \ldots, \bm{\theta}_{K-1}, \bm{\kappa}, \overline{\bm{\theta}}^A_0, \ldots, \overline{\bm{\theta}}^A_{K-1}, \overline{\bm{\kappa}}^A) \\
 &= \bra{\psi}\hat{U}^\dag(\bm{\theta}_A)\hat{\mc{H}}(\bm{x},\bm{\kappa})\hat{U}(\bm{\theta}_A)\ket{\psi} \\
 & + \sum_{S=0}^{K-1}\sum_i \overline{\theta}^A_{Si} \pdv{F^{\mr{VQD}}_S}{\theta_{Si}}  + \sum_{p>q} \overline{\kappa}^A_{pq}\frac{\partial E^{\mr{SA}}}{\partial\kappa_{pq}},
\end{split}
\end{equation}
where $\theta_{Si}$ is $i$th element of $\bm{\theta}_S$, and $\overline{\theta}^A_{Si}, \overline{\kappa}^A_{pq}$ are Lagrange multipliers.
We impose the extremal condition for all parameters except for $\bm{x}$ on the Lagrangian,
\begin{equation}\label{eq:lagrangian extremal vqd}
\begin{split}
 \pdv{L_A^\mr{VQD}}{\theta_{Tj}} 
 = \pdv{L_A^\mr{VQD}}{\kappa_{mn}}
 = \pdv{L_A^\mr{VQD}}{\overline{\theta}_{Tj}^A}
 = \pdv{L_A^\mr{VQD}}{\overline{\kappa}_{mn}^A} = 0
 \end{split}.
\end{equation}
The last two equations in Eq.~\eqref{eq:lagrangian extremal vqd} are satisfied when SA-OO-VQD converges, $\bmkp = \bmkp^*, \bmth_S = \bmth_S^*$.
The first two equations in Eq.~\eqref{eq:lagrangian extremal vqd} can be formulated as a linear equation,
\begin{equation}\label{eq:linear equation vqd}
 \begin{pmatrix}
  \mathbb{H}^{\mr{VV}}&\mathbb{H}^{\mr{VO}} \\
  \mathbb{H}^{\mr{OV}} & \mathbb{H}^{\mr{OO}}
 \end{pmatrix}
 \begin{pmatrix}
  \overline{\bm{\theta}}^A_0 \\ \vdots \\ \overline{\bm{\theta}}^A_{K-1} \\ \overline{\bm{\kappa}}^A
 \end{pmatrix}
 =-\begin{pmatrix} \bm{0} \\ \vdots \\ \bm{0} \\\bm{g}^A \end{pmatrix},
\end{equation}
where we define
\begin{equation}\label{eq:def of z-vect}
\begin{split}
 &g^A_{(mn)}=\frac{\partial E_A}{\partial\kappa_{mn}}, \\
 &\mathbb{H}^{\mr{VV}}_{(Tj)(Si)}=\frac{\partial^2F^{\mr{VQD}}_S}{\partial\theta_{Si}\partial\theta_{Tj}},  \mathbb{H}^{\mr{VO}}_{(Tj)(pq)}=\frac{\partial^2E^{\mr{SA}}}{\partial\kappa_{pq}\partial\theta_{Tj}}\\
  &\mathbb{H}^{\mr{OV}}_{(mn)(Tj)}=\frac{\partial^2F^{\mr{VQD}}_S}{\partial\theta_{Tj}\partial\kappa_{mn}}, \mathbb{H}^{\mr{OO}}_{(mn)(pq)}=\frac{\partial^2E^{\mr{SA}}}{\partial\kappa_{pq}\partial\kappa_{mn}}, \\
 &E_A(\bm{x},\bmkp, \bmth_A) = \bra{\psi}\hat{U}^\dag(\bm{\theta}_A)\hat{\mc{H}}(\bm{x},\bm{\kappa})\hat{U}(\bm{\theta}_A)\ket{\psi}, \\
 &E^{\mr{SA}}(\bm{x}, \bmkp, \bmth_0, \ldots, \bmth_{K-1}) = \sum_{S=0}^{K-1} \omega_S^{\mr{SA}} E_S(\bm{x}, \bmkp, \bmth_S).  
 \end{split}
\end{equation}
Note that we use $\pdv{E_A}{\theta_{Tj}}=0$ for $T\neq A$ and $\pdv{E_A}{\theta_{Aj}} = 0$ for the optimal $\bm{\theta}_A^*$.
All the elements in the linear equation can be determined by standard expectation value measurements by quantum computers and the 1,2-RDMs evaluated in SA-OO-VQD~\cite{Mizukami2020}.
We can compute the values of the Lagrange multipliers by solving the equation with classical computers.
The details of the evaluation and its computational cost are explained in SI.

Once the values of the Lagrange multipliers $\overline{\bm{\theta}}^A_0,\cdots\overline{\bm{\theta}}^A_{K-1}, \overline{\bm{\kappa}}^A$ are determined by solving the linear equation Eq.~\eqref{eq:linear equation vqd} (which we denote $\overline{\bm{\theta}}_0^{A*},\cdots\overline{\bm{\theta}}_{K-1}^{A*},\overline{\bm{\kappa}}^{A*}$), the analytical gradient of the energy $E_A$ with respect to $x_\mu$ is easily computed by leveraging the extremal conditions of the Lagrangian~\eqref{eq:lagrangian extremal vqd}: That is,
\begin{strip}
\begin{equation}\label{eq:dE/dx in VQD}
 \begin{aligned}
  &\frac{dE_A^*(\bm{x})}{dx_\mu} = \bra{\psi}\hat{U}^\dag(\bm{\theta}_A^*)\frac{\partial\hat{\mc{H}}(\bm{x},\bm{\kappa}^*)}{\partial x_\mu}\hat{U}(\bm{\theta}_A)\ket{\psi} + \sum_{S=0}^{K-1} \sum_i \overline{\theta}_{Si}^{A*}\frac{\partial}{\partial\theta_{Si}}\bra{\psi}\hat{U}^\dag(\bm{\theta}_S^*)\frac{\partial\hat{\mc{H}}(\bm{x},\bm{\kappa}^*)}{\partial x_\mu}\hat{U}(\bm{\theta}_S^*)\ket{\psi} \\
  &+ \sum_{S=0}^{K-1} \sum_{p>q} \overline{\kappa}_{pq}^{A*} \omega_S^{\mr{SA}}\frac{\partial}{\partial\kappa_{pq}}\bra{\psi}\hat{U}^\dag(\bm{\theta}_S^*)\frac{\partial\hat{\mc{H}}(\bm{x},\bm{\kappa}^*)}{\partial x_\mu}\hat{U}(\bm{\theta}_S^*)\ket{\psi}.
 \end{aligned}
\end{equation}
\end{strip}
Again, all terms in the above equation can be calculated by utilizing the 1,2-RDMs and measuring several expectation values on quantum computers.
As we show in SI, apart from the cost to perform the optimization in SA-OO-VQD and obtain its (eigen)energies, the additional number of distinct quantum circuits to be measured to evaluate the analytical derivative is $\order{KM^2 N^4}$, where $M$ is the number of elements in the circuit parameters $\bmth$ and $N$ is the number of qubits.
Then a very naive (and possibly worst) estimate for the required number of measurements to evaluate the derivative with precision $\epsilon$ is $\order{K^2M^4 N^8/\epsilon^2}$ because measuring outcomes of quantum circuits with precision $\epsilon'$ takes $\order{1/\epsilon'^2}$ measurements and we assume each outcome of a circuit should be measured with precision $\epsilon' = \epsilon/(KM^2 N^4)$.
\section{Numerical Experiment}
As an application of our theory to photochemical reactions, we have performed calculations of the {\it cis-trans} photoisomerization reaction of the {\it cis}-TFP molecule, shown in Scheme~\ref{Sch:TFP}, including its minimum energy conical intersection (CI$_{\rm MIN}$).
Note that benchmarking of {\it cis-trans} isomerizations has typically been performed with ethylene and penta-2,4-dieniminium cation~\cite{gozem2013}.
\begin{scheme}
\centering
\includegraphics[width=1.0\columnwidth]{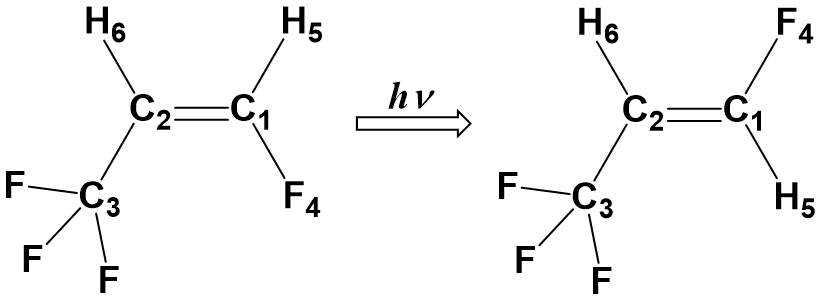}
\caption{{\it cis-trans} photoisomerization reaction of {\it cis}-TFP (with atom numbering).}
\label{Sch:TFP}
\end{scheme}

In the static view, the {\it cis-trans} photoisomerization reaction of {\it cis}-TFP is considered to proceed in the following three phases:
(1) 
After photoexcitation to the S$_\textrm{1-cis}$ Franck-Condon (S$_\textrm{1-cis-FC}$) state, the molecule goes down to the S$_1$ minimum (S$_\textrm{1-MIN}$) point without any barriers.
(2)
The molecule reaches the S$_1$/S$_0$ CI$_{\rm MIN}$, which is located spatially and energetically near S$_\textrm{1-MIN}$. 
(3) The nonradiative decay from S$_1$ to S$_0$ through the CI$_{\rm MIN}$ occurs with branching backward to the {\it cis}-TFP isomer or forward to the {\it trans}-TFP one.
We chose TFP as one of the simplest molecules exhibiting photoisomerization reaction pathways explained above.

We numerically performed a search and characterization of the CI$_{\rm MIN}$ and determination of the minimun energy path (MEP) regarding the {\it cis-trans} photoisomerization reaction of {\it cis}-TFP. We compare two results obtained by our theory for the gradient of energy, i.e., gradient of $\hat{\mc{H}}(\bm{x},\bm{\kappa}^*(\bm{x}))$ as shown in Eq.~\eqref{eq:dE/dx in VQD}, and by the conventional approach with classical computers.
We call the former (latter) as quantum (classical) approach.

In particular, we calculated three lowest singlet states (S$_0$, S$_1$, and S$_2$) using SA-OO-VQD for the quantum approach and its classical equivalent, SA-CASSCF, for the classical approach.
In both approaches, the active space consists of two orbitals (HOMO and LUMO) with two electrons, and the 6-31G basis set was employed.
The weights for SA for three states were taken identical ($\omega^\mr{SA}_0 = \omega^\mr{SA}_1 = \omega^\mr{SA}_2$ = 1/3).
The SA-OO-VQD energy ($E_A(\bm{x},\bm{\kappa}^*,\bm{\theta}_A^*)$ in Eq.~\eqref{eq:def of z-vect}) and its analytical derivative ($dE_A^*(\bm{x})/dx_\mu$ in Eq.~\eqref{eq:dE/dx in VQD}) were obtained by simulating quantum circuits and measurements on classical computers assuming that there are no error, noise and statistical fluctuation in the output of quantum computers.
The $\mr{QAMUY}^\mr{TM}$ software developed by QunaSys Inc. was used to compute those SA-OO-VQD energies and gradients.
Further details including the ansatz employed for VQD and the hyperparameters $\beta_T$ are explained in SI. 
The classical SA-CASSCF calculations were performed by Molpro2015~\cite{MOLPRO-WIREs,MOLPRO2020,MOLPRO}. 

We first show the results of the optimization and characterization of the S$_1$/S$_0$ CI$_{\rm MIN}$ structure.
The S$_1$/S$_0$ CI$_{\rm MIN}$ optimization, starting from the S$_\textrm{1-MIN}$ point obtained by Molpro2015, was performed 
using constraint energy minimization by the updated branching plane (BP) approach~\cite{ConIn_Maeda2010}, which requires a gradient of each eigenenergy we have formulated above, in the quantum approach.
In the classical approach, 
the CI$_{\rm MIN}$ was calculated using the gradient difference (GD) and the derivative coupling (DC) by Molpro2015. The optimized CI$_{\rm MIN}$ was characterized by two BP vectors (denoted by $\bm{g}$ and $\bm{h}$) and four conical parameters ($s_x$, $s_y$, $d_{gh}$, and  and $\Delta_{gh}$)~\cite{Conical_YarkonyJPCA2001}.
The BP vectors are obtained by orthogonalizing the GD and DC vectors by utilizing a unitary transformation of the two degenerate states.

The four conical parameters can perturbatively describe two crossing state energies in the vicinity of CI${}_\mr{MIN}$ in the BP as
\begin{equation}
\begin{split}
E_{\pm}(\rho,\varphi)&=
\rho\Big(s_x\cos\varphi + s_y\sin\varphi\\& 
\pm d_{gh} \sqrt{1+\Delta_{gh}\cos2\varphi} \Bigr),
\end{split}
\label{eq:E_BP}
\end{equation}
where $\rho$ is a radius and $\varphi$ is an angle in the polar coordinate centered at the CI$_{\rm MIN}$ and $\pm$ denotes two crossing state energies around it.
Computational details to determine the BP vectors and the conical parameters are described in SI.

\begin{figure*}[!htb]
\centering
\subfloat[][Quantum]{
\includegraphics[width=0.38\textwidth]{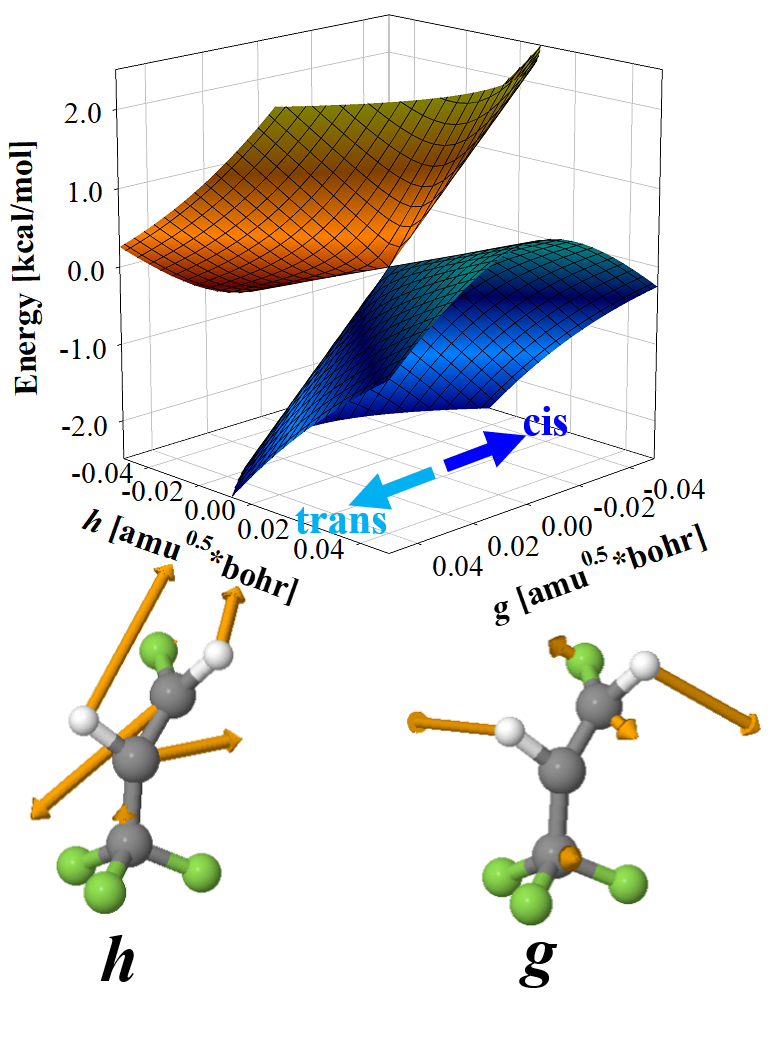}
\label{fig:BP_Q}}
\qquad
\subfloat[][Classical]{
\includegraphics[width=0.38\textwidth]{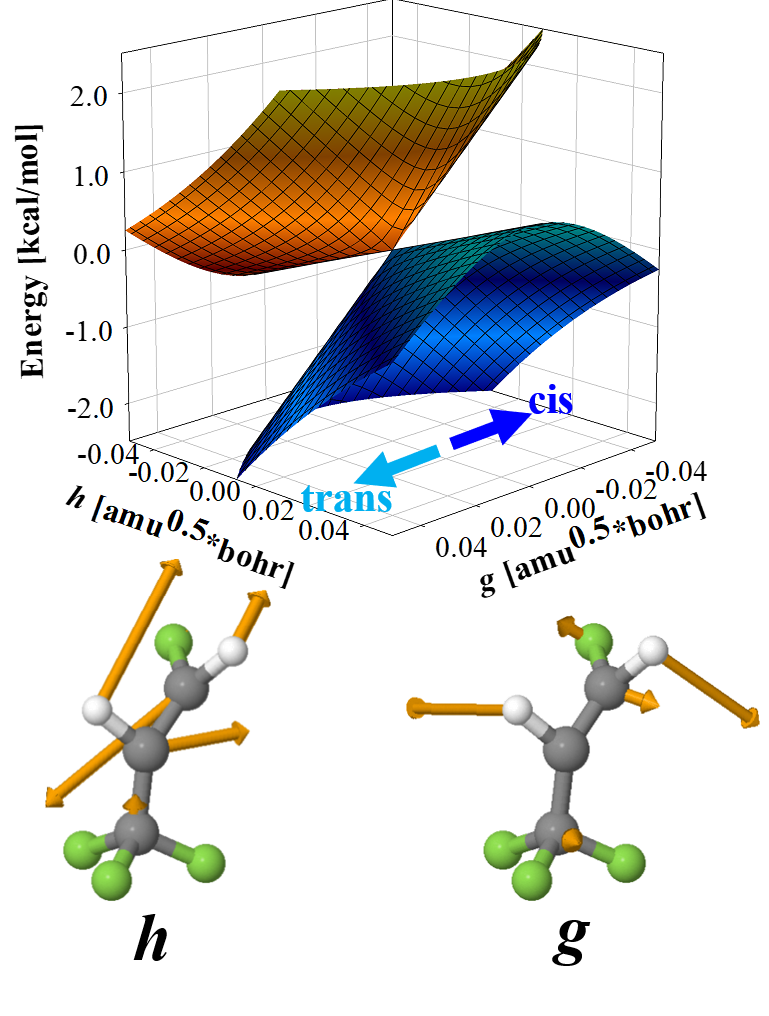}
\label{fig:BP_C}}
\captionsetup{format=plain,font=normalsize}
\caption{(Upper) S$_1$/S$_0$ energies centered at the CI$_{\rm MIN}$ in the BP and (lower) the BP vectors of TFP 
obtained by the quantum (a) and classical (b) approaches. The BP vectors are depicted in the mass-weighted coordinates.
}
\label{fig:BP}
\end{figure*}

Tables \ref{Tab:CI_E}, \ref{Tab:CI_xyz} and Figure~\ref{fig:BP} summarize the optimized S$_1$/S$_0$ CI$_{\rm MIN}$ and its characteristics.
The S$_1$/S$_0$ energies and the parameters for S$_1$/S$_0$ CI$_{\rm MIN}$ obtained by the quantum approach are in an excellent agreement with those by the classical approach.
This means that SA-OO-VQD and its analytical energy derivative are accurate enough to locate conical intersections and hence elucidate photochemical reactions at the same level as the classical counterpart (SA-CASSCF).

\begin{table}[bt]
\centering 
 \caption{S$_1$/S$_0$ energies at the optimized CI$_{\rm MIN}$ of TFP. \label{Tab:CI_E}} 
\begin{tabular}{@{\extracolsep{2pt}} cccc} 
\\[-1.8ex]\hline 
\hline \\[-1.8ex] 
Method & $E({\rm S_0})$ & $E({\rm S_1})$ & $\Delta E$ \\ 
& [Hartree] & [Hartree] & [kcal/mol] \\
\hline \\[-1.8ex] 
Quantum & -512.144700 & -512.144679 & 0.014 \\ 
Classical & -512.144694 & -512.144679 & 0.009 \\ 
\hline \\[-1.8ex] 
\end{tabular} 
\end{table} 

\begin{table*}
\centering 
 \caption{Structural and conical parameters at the optimized S$_1$/S$_0$ CI$_{\rm MIN}$ of TFP.} 
  \label{Tab:CI_xyz} 
\begin{tabular}{cccccccc} 
\\[-1.8ex]\hline 
\hline \\[-1.8ex] 
Method & $R({\rm C_1C_2})$ & $\theta({\rm C_3C_2H_6})$ & $\phi({\rm H_5C_1C_2H_6})$ & $s_x$ & $s_y$ & $d_{gh}$ & $\Delta_{gh}$ \\ 
& [\AA] & [degrees] & [degrees] \\
\hline \\[-1.8ex] 
Quantum & 1.284 & 123.7 & 88.8 & -0.0023 & 0.0707 & 0.0574 & 0.3530\\ 
Classical            & 1.284 & 123.8 & 88.7 & -0.0023 & 0.0707 & 0.0574 & 0.3514\\ 
\hline \\[-1.8ex] 
\end{tabular} 
\end{table*} 

Next, we show the result of MEP determination of {\it cis-trans} photoisomerization of {\it cis}-TFP.
Corresponding to the three phases of the photochemical reaction previously mentioned in this section, 
the calculations of the MEP were divided as (1) S$_\textrm{1-cis-FC}$ $\to$ S$_\textrm{1-MIN}$, (2) S$_\textrm{1-MIN}$ $\to$ S$_1$/S$_0$ CI$_{\rm MIN}$, and (3)
S$_1$/S$_0$ CI$_{\rm MIN}$ $\to$ S$_\textrm{0-cis}$ or S$_\textrm{0-trans}$.
The MEP calculations were performed by the Gonzalez-Schlegel method~\cite{Gonzalez1990} both in the quantum and classical approach.
As starting points of the first and the second parts of the MEP, we used the S$_\textrm{0-cis}$ and S$_\textrm{1-MIN}$ structures optimized by Molpro2015, respectively. 
More computational details are described in SI.

\begin{figure*}[!htb]
\centering
\subfloat[][Quantum]{
\includegraphics[width=0.45\textwidth]{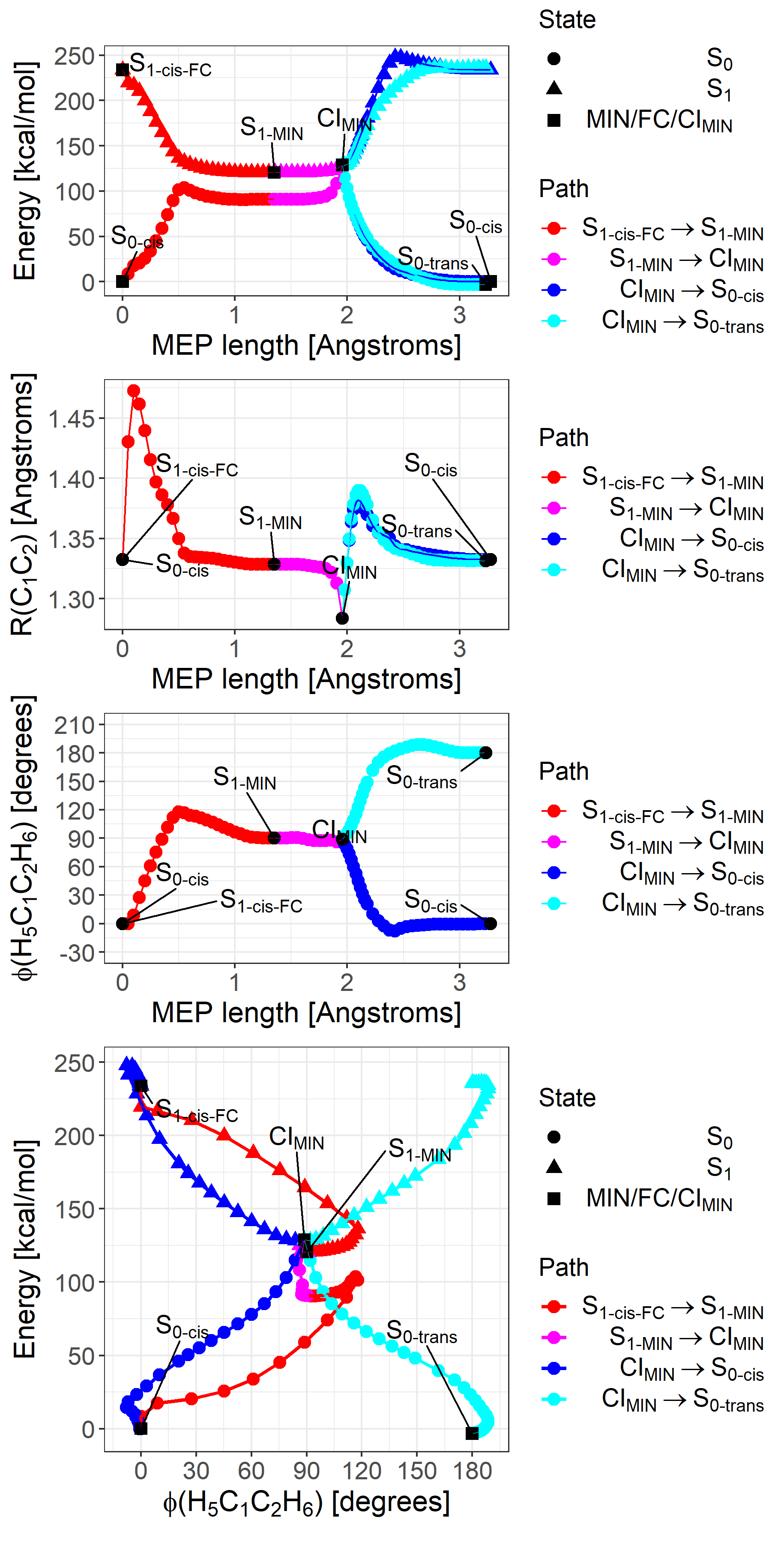}
\label{fig:MEP-PATH_Q}}
\qquad
\subfloat[][Classical]{
\includegraphics[width=0.45\textwidth]{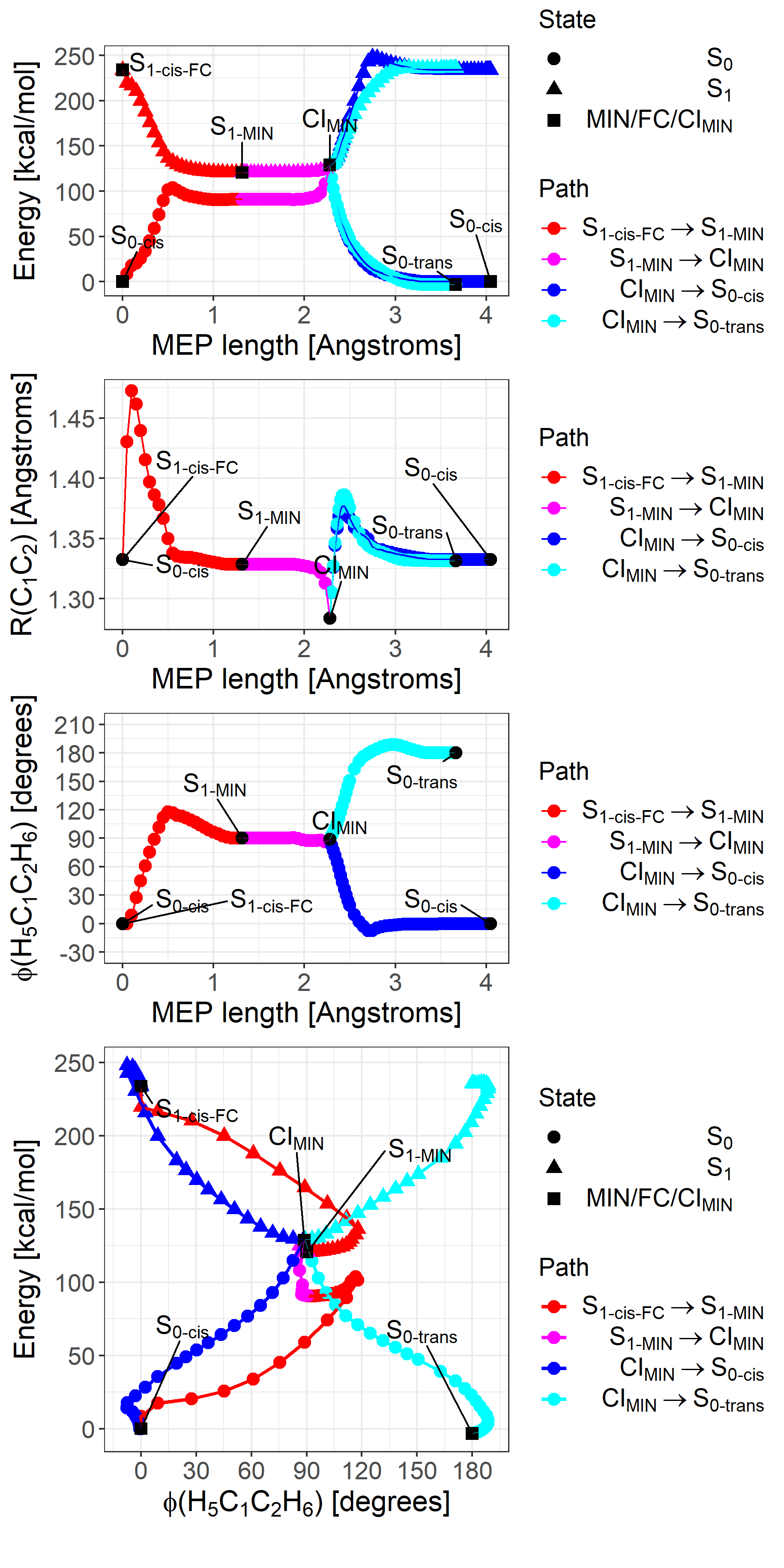}
\label{fig:MEP-PATH_C}}

\captionsetup{format=plain,font=normalsize}
\caption{
(From top to third rows) The S$_1$/S$_0$ energies, C$_1$C$_2$ bond length ($R$(\textrm{C$_1$C$_2$})), 
and H$_5$C$_1$C$_2$H$_6$ dihedral angle ($\phi$(\textrm{H$_5$C$_1$C$_2$H$_6$})) along the MEP length and 
(bottom) the S$_1$/S$_0$ energies and $\phi$(\textrm{H$_5$C$_1$C$_2$H$_6$}) in the MEP of the {\it cis-trans} photoisomerization reaction of {\it cis}-TFP obtained by the quantum (a) and classical (b) approaches.
}
\label{fig:MEP-PATH}
\end{figure*}

Figure~\ref{fig:MEP-PATH} shows the S$_1$/S$_0$ energies and two structural parameters of C$_1$C$_2$ bond length ($R$(\textrm{C$_1$C$_2$})) and H$_5$C$_1$C$_2$H$_6$ dihedral angle ($\phi$(\textrm{H$_5$C$_1$C$_2$H$_6$})), which characterize the {\it cis-trans} photoisomerization reaction, along the MEP length.
We also show the S$_1$/S$_0$ energies versus the dihedral angle $\phi$(\textrm{H$_5$C$_1$C$_2$H$_6$}) in the MEP. 
Again, it can also be seen from Figure~\ref{fig:MEP-PATH} that the resulting MEPs by the quantum approach are quite similar to those by the classical one. 
Furthermore, it should be noted that the entire energy curve along the MEP obtained by the SA-OO-VQE approach is totally smooth. 
This further confirms that the {\it cis-trans} photoisomerization reaction of TFP can be analyzed by the quantum approach.

\section{Conclusion}
In this work, we present a practical method to calculate the analytical energy derivative of SA-MC calculations by quantum computers and discuss its possibility for the analysis of photochemical reactions.
Our method leverages the Lagrangian method:
the multipliers are determined by classically solving linear equations whose components are evaluated by quantum computers, and the values of the multipliers are substituted into the expression of the analytical derivative of the energies. 
Numerical experiment emulating quantum computers by classical computers shows that energy derivatives obtained by our method can be leveraged to analyze {\it cis-trans} photoisomerization of TFP molecule with the same accuracy as the classical counterpart, SA-CASSCF.
Although our simulations in this work as well as currently-available quantum computers are not large enough to rival classical ones, our proposed method can be utilized when one tackles large problems (molecules) with quantum computers in the near future.
We stress that the use of quantum computers (SA-OO-VQE) will be important to replace SA-CASSCF.

As explained in Introduction, one has to use the perturbation theory in ``classical" SA-CASSCF calculations to take the dynamical electron correlation into account since the size of the active space is limited to a few dozen.
We expect that quantum computers will eventually be able to handle an active space of several hundred orbitals. 
If this happens, it will be possible to simultaneously treat both dynamical and static electron correlations in the framework of SA-OO-VQE, allowing us to analyze photochemical reactions in a more black-box and robust manner.
This work shows the first step for exploiting quantum computers along this direction by providing concrete procedures to calculate a quantity central to simulate photochemical reactions.

Interesting future directions of our work include the development of the method to calculate the non-adiabatic couplings, which is vital to investigate the dynamic view of photochemical reactions as well as the static view of them tested in this study.

\begin{acknowledgement}
 A part of this work was performed for Council for Science, Technology and Innovation (CSTI), Cross-ministerial Strategic Innovation Promotion Program (SIP), ``Photonics and Quantum Technology for Society 5.0" (Funding agency: QST).
 This work was supported by MEXT Quantum Leap Flagship Program (MEXT Q-LEAP) Grant Number JPMXS0118067394 and JPMXS0120319794.
 W.M. wishes to thank Japan Society for the Promotion of Science (JSPS) KAKENHI No.\ 18K14181 and JST PRESTO No.\ JPMJPR191A.
 We also acknowledge support from JST COI-NEXT program.
\end{acknowledgement}

 \clearpage
 \appendix
 \begin{suppinfo}
\renewcommand{\theequation}{S\arabic{equation}} 
\setcounter{equation}{0} 

\section{Quantum chemistry calculations by quantum computers}
Quantum computers~\cite{Feynman1982} are potentially capable of solving problems that classical computers cannot answer in a reasonable amount of time.
This potential has been the driving force behind recent active researches on quantum computers and efforts to make them a reality. 
As a result of these efforts, relatively small-scale quantum computers called noisy intermediate-scale quantum devices (NISQs) are being developed~\cite{Preskill2018}.
NISQs are without error tolerance and have a short coherent time, so they cannot execute complicated quantum circuits (computations) such as the celebrated quantum phase estimation to find eigenspectrum of sparse matrix~\cite{kitaev1995quantum, Cleve1998, AspuruGuzik2005}.
However, they still have potentially large computational powers; it has been shown that some of them are already faster than the most advanced supercomputers for some specific tasks~\cite{Arute2019,Zhong1460}.

One of the most promising applications of NISQs is found in quantum chemistry.
Especially, it is expected that NISQs are utilized in solving electronic states of molecules under the Born-Oppenheimer approximation, i.e., solving eigenspectrum of Hamiltonian like Eq.~(1).
When the number of molecular orbitals (MOs) is $\tilde{N}$, the size of the matrix representation of the Hamiltonian~(1) 
is $\order{e^{\tilde{N}}}$.
This fact prevents us from solving the Hamiltonian of a large molecule by classical computers.
On the other hand, quantum computers can handle the Hamiltonian with $N=2\tilde{N}$ qubits.
The linear scaling of the number of qubits with respect to $\tilde{N}$ is a distinctive advantage of quantum computers.
See Refs.~\citenum{mcardle2018quantum, Cao2018} for a review of quantum chemistry calculations with quantum computers.

\section{Review of SA-OO-SSVQE and SA-OO-MCVQE}

\begin{figure}[tb]
    \centering
    \includegraphics[width=.48\textwidth, trim=20 20 20 20]{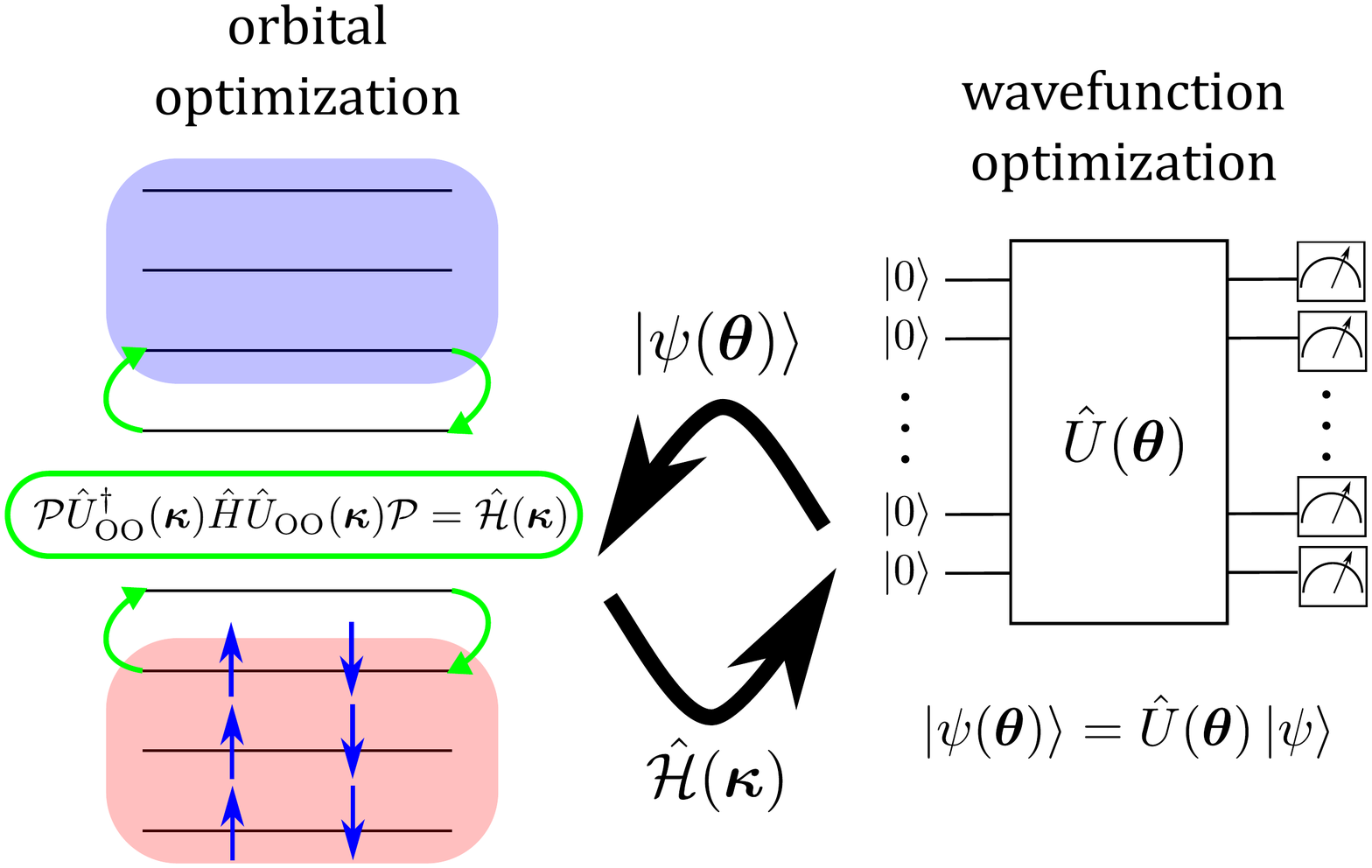}
    \caption{Schematic picture of the SA-OO VQEs.
    \label{fig:sa-oo-vqe_repre}}
\end{figure} 

In the main text, we treat state-averaged (SA) orbital-optimized (OO) variational quantum deflation (VQD), shortly SA-OO-VQD, as an example of
SA-OO variational quantum eigensolvers (SA-OO-VQEs) and describe the formula of analytical derivative of energies.
In the following two sections, we provide formulas for the analytical derivative of energies for SA-OO subspace-search VQE (SA-OO-SSVQE) and SA-OO multistate-contracted VQE (SA-OO-MCVQE).
In this section, we review SA-OO-SSVQE and SA-OO-MCVQE.

In SA-OO-VQEs, or SA-MC calculations with VQEs by quantum computers, the algorithms consist of two optimization steps and continue them until the SA energy $E^{\mr{SA}} = \sum_{S=0}^{K-1} \omega_S^{\mr{SA}} E_S$ converges, as depicted in Fig.~\ref{fig:sa-oo-vqe_repre}.
The first step is to optimize the trial wavefunction (or, ansatz state) to obtain eigenstates of the active space Hamiltonian $\hat{\mathcal{H}}(\bm{x}, \bm{\kappa})$ with fixed orbital parameters $\bmkp$ (left panel).
The second step is to update $\bmkp$ to lower the SA energy, which is typically done by using the 1,2-RDMs of the optimized wavefunction obtained in the first step.

\subsection{SA-OO-SSVQE\label{subsubsec:sa-oo-vqe}}
We first summarize SSVQE~\cite{nakanishi2018subspace} briefly.
For a given the Hamiltonian $\hat{\mathcal{H}}$, we prepare mutually orthogonal input states $\left\{\ket{\psi_S}\right\}_{S=0}^{K-1}$, and construct an ansatz circuit described by some unitary operator $\hat{U}(\bm{\theta})$ with circuit parameters $\bm{\theta}$. The parameters $\bm{\theta}$ are optimized to minimize the following cost function,
\begin{equation}
    E^{\mr{SSVQE}}(\bm{\theta})=\sum_{S=0}^{K-1} \omega_S^\mr{VQE} \bra{\psi_S}\hat{U}^\dag(\bm{\theta})\hat{\mc{H}}\hat{U}(\bm{\theta})\ket{\psi_S},
    \label{Seq: SSVQE cost}
\end{equation}
where $\omega_S^\mr{VQE}$ is a weight that satisfies $\omega_0^\mr{VQE}>\cdots>\omega_{K-1}^\mr{VQE}>0$.
When the global minimum of the cost function is reached at $\bm{\theta}^*$, the lowest $K$ eigenstates are written as $\ket{\psi_S(\bm{\theta}^*)} = \hat{U}(\bm{\theta}^*)\ket{\psi_S} (S=0\sim K-1)$ and the associated eigenenergies are $\braket{\psi_S(\bm{\theta}^*)|\hat{\mathcal{H}}|\psi_S(\bm{\theta}^*)}.$
We note that the weight for the SSVQE, $\omega_S^\mr{VQE}$, is not related to that for the SA energy, $\omega_S^{\mr{SA}}$.

In SA-OO-SSVQE, We prepare mutually orthogonal input states $\{\ket{\psi_S}\}_{S=0}^{K-1}$ and an ansatz circuit $\hat{U}(\bm{\theta})$ in the same way as SSVQE.
We optimize the circuit parameters $\bmth$ so as to minimize the cost function [Eq.~\eqref{Seq: SSVQE cost}] for the active space Hamiltonian $\hat{\mc{H}}(\bm{x}, \bm{\kappa})$.
Note that this optimization is done with keeping the orbital parameters $\bm{\kappa}$ fixed.
After obtaining the eigenstates and energies of $\hat{\mc{H}}(\bm{x}, \bm{\kappa})$ as a result of SSVQE subroutine, we optimize $\bmkp$.
The state-averaged energy $E^{\mr{SA}}$, expressed as 
\begin{equation}\label{eq:sa-oo}
 E^{\mr{SA}}(\bm{\kappa})=\sum_{S=0}^{K-1} \omega_S^{\mr{SA}}\bra{\psi_S}\hat{U}^\dag(\bmth^*)\hat{\mathcal{H}}(\bm{x}, \bm{\kappa})\hat{U}(\bmth^*)\ket{\psi_S}
\end{equation}
in this case, is optimized with respect to $\bm{\kappa}$ with the optimized circuit parameters $\bmth^*$ determined in the previous step.
This optimization is performed with the Newton-Raphson method by using  $\pdv{E^{\mr{SA}}}{\bmkp}, \pdv{E^{\mr{SA}}}{\bmkp}{\bmkp}$, which can be evaluated with 1,2-RDMs of $\{\ket{\psi_S(\bmth^*)}\}_S$ (see ``Evaluation of derivatives appearing in the formulas" section).
The two procedures are repeated until the state-averaged energy $E^{\mr{SA}}(\bm{\kappa})$ converges.

\subsection{SA-OO-MCVQE \label{subsec: SA-OO-MCVQE review}}
MCVQE~\cite{Parrish2019PRL} also works as an eigensolver for a given Hamiltonian $\hat{\mc{H}}$ based on the variational principle of quantum mechanics.
It consists of two steps.
In the first step, circuit parameters $\bm{\theta}$ of the ansatz quantum circuit $\hat{U}(\bmth)$ with the orthonormal initial states $\left\{\ket{\psi_S}\right\}_{S=0}^{K-1}$ are optimized by minimizing the following cost function,
\begin{equation}\label{Seq: average energy}
    E^{\mr{MCVQE}}(\bm{\theta})=\frac{1}{K}\sum_{S=0}^{K-1} \bra{\psi_S}\hat{U}^\dag(\bm{\theta})\hat{\mc{H}}\hat{U}(\bm{\theta})\ket{\psi_S}.
\end{equation}
In the second step, we diagonalize the Hamiltonian within the subspace spanned by $\{\ket{\psi_S(\bm{\theta}^*)}=\hat{U}(\bm{\theta}^*)\ket{\psi_S}\}_S$, where $\bmth^*$ is the optimal parameters.
Namely, the Hamiltonian in the subspace is represented by the $K\times K$ matrix $h$ whose matrix elements are given by
\begin{equation}
    h_{ST} = \bra{\psi_S}\hat{U}^\dag(\bm{\theta}^*)\hat{\mc{H}}\hat{U}(\bm{\theta}^*)\ket{\psi_T}.
\end{equation}
These elements $h_{ST}$ can be evaluated by quantum computers if we prepare the superposed states like $1/\sqrt{2}(\ket{\psi_S}\pm\ket{\psi_T}), 1/\sqrt{2}(\ket{\psi_S}\pm i\ket{\psi_T}) $~\cite{Parrish2019PRL} and diagonalization of $h$ is performed by classical computers.
We denote $A$th (classical) eigenvector of $h$ as $v^{(A)}$, which satisfies for any $S$,
\begin{equation}\label{Seq: mcvqe eigeneqn}
    \sum_{T=0}^{K-1} h_{ST} \, v^{(A)}_T = E_A v^{(A)}_S,
\end{equation}
where $E_A$ is the corresponding eigenvalue.
Equation~\eqref{Seq: mcvqe eigeneqn} implies that $A$th excited state $\ket{\Psi_A}$ of the Hamiltonian $\hat{\mc{H}}$ is expressed as
\begin{equation} \label{Seq: mcvqe wfn}
    \ket{\Psi_A}=\sum_{S=0}^{K-1} v^{(A)}_S \hat{U}(\bm{\theta}^*)\ket{\psi_S}
\end{equation}
with the energy $E_A$.

In SA-OO-MCVQE, given by the active space Hamiltonian $\hat{\mc{H}}(\bm{x}, \bmkp)$ 
for some fixed $\bm{\kappa}$, we optimize the circuit parameter $\bmth$ and obtain classical vectors $v^{(A)}$ as explained above.
After that, the orbital parameter $\bm{\kappa}$ is updated so that the SA energy $E^{\mr{SA}}$, expressed as 
\begin{align*}
 &E^{\mr{SA}}(\bmkp) = \\
 &\sum_{A,S,T=0}^{K-1} \omega_S^{\mr{SA}} v_S^{(A)*} v_T^{(A)} \bra{\psi_S}\hat{U}^\dag(\bmth^*)\hat{\mc{H}}(\bm{x}, \bmkp)\hat{U}(\bmth^*)\ket{\psi_T}  
\end{align*}
in this case, gets small.
We repeat the updates of $\bmth, v^{(A)}$ and $\bmkp$ until the SA energy converges.

\section{Derivation of Gradient for SA-OO-SSVQE and SA-OO-MCVQE}
The formulas for the analytical derivatives of SA-OO-SSVQE and SA-OO-MCVQE energies are also based on the Lagrangian method, as the same in SA-OO-VQD described in the main text.
We can summerize the general scheme for computing the energy gradient of $A(=0,\ldots,K-1)$th eigenstate in SA-OO-VQEs as follows (Fig.~\ref{fig:analytical_grad}):
\begin{enumerate}
    \item[0.] Perform SA-OO-VQD, SA-OO-SSVQE, or SA-OO-MCVQE and obtain the optimal orbital parameters $\bm{\kappa}^*$, the optimal circuit parameters $\bm{\theta}^*$, and classical vector $v^{(A)}$ (only for SA-OO-MCVQE). 
    \item Define a Lagrangian $L_A(\bm{x}, \bm{\theta}, \bm{\kappa}, \overline{\bm{\theta}}^A, \overline{\bm{\kappa}}^A)$ and impose the extremal condition on $L_A$ with respect to the parameters other than $\bm{x}$. Here, $\overline{\bmth}^A$ and $\overline{\bm{\kappa}}^A$ are Lagrange multipliers.
    \item Calculate the multipliers satisfying the extremal condition by solving a linear equation
    \begin{equation}\label{eq:linear multipliers}
        \begin{pmatrix}
        \mathbb{H}^{\mr{VV}} & \mathbb{H}^{\mr{VO}}\\\mathbb{H}^{\mr{OV}} & \mathbb{H}^{\mr{OO}}
        \end{pmatrix}
        \begin{pmatrix}
        \overline{\bm{\theta}}^A\\\overline{\bm{\kappa}}^A
        \end{pmatrix}=-\begin{pmatrix}
        \bm{f}^A\\\bm{g}^A
        \end{pmatrix},
    \end{equation}
    where $\mathbb{H}^{XY}$ ($X,Y=V,O$, denoting the derivative for $\bmth$ and $\bmkp$, respectively) is a Hessian matrix of $L_A$, and $\bm{f}^A, \bm{g}^A$ is a first-order derivative of the $A$th eigenenergy.
    The concrete forms of those quantities and the way to determine it on quantum computers are explained later. 
    \item By combining the values of multipliers and several quantities that are easy to evaluate on quantum computers, analytical gradients can be calculated.
    The concrete forms of the final result slightly vary among VQD, SSVQE, and MCVQE.
\end{enumerate}

\begin{figure}[tb]
    \centering
    \includegraphics[width=.5\textwidth]{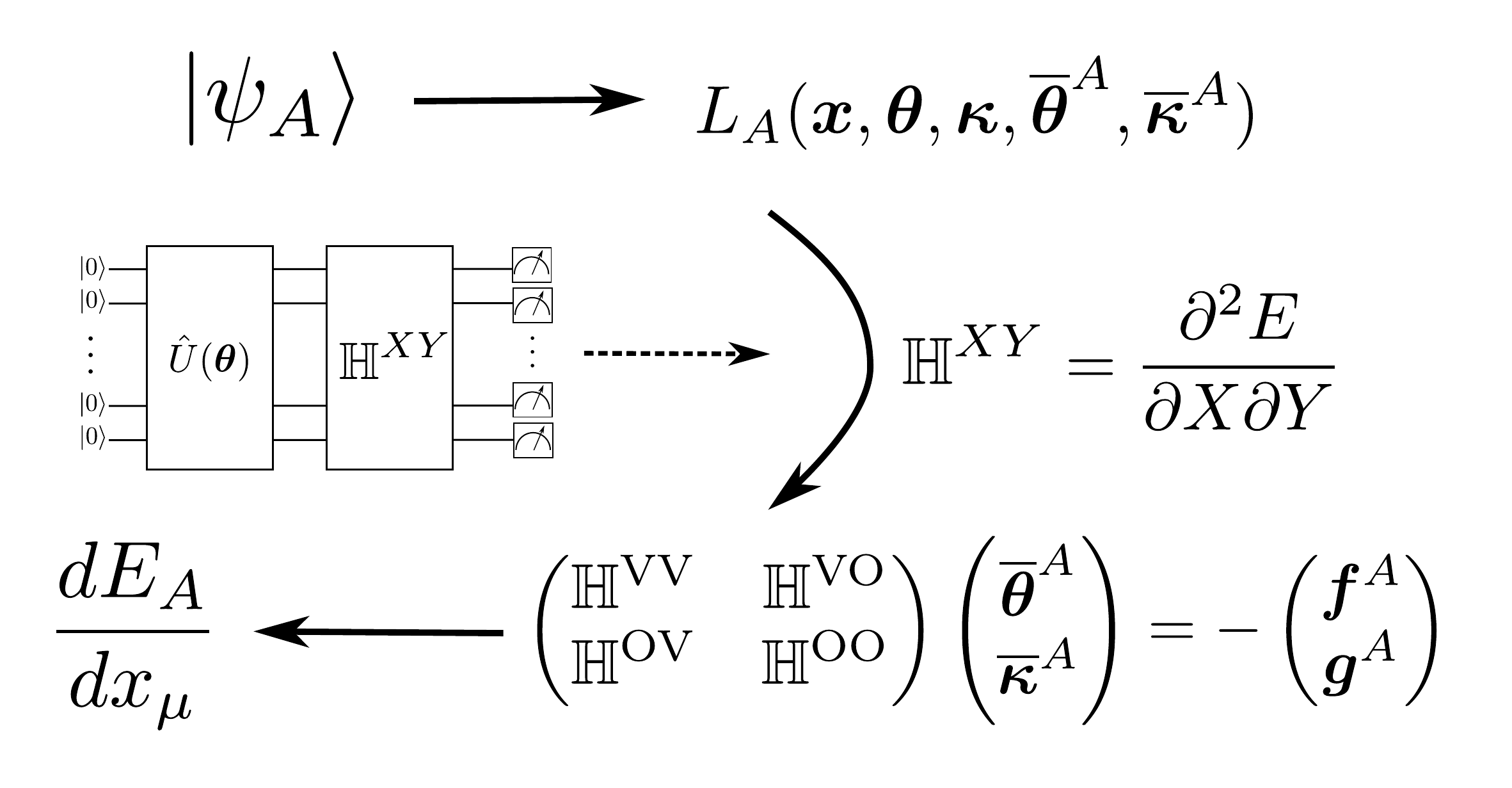}
    \caption{A schematic diagram of our procedure to calculate the energy derivative: (i) define the Lagrangian $L_A(\bm{x}, \bm{\theta}, \bm{\kappa}, \overline{\bm{\theta}}^A, \overline{\bm{\kappa}}^A)$ with the Lagrange multipliers $\overline{\bm{\theta}}^A, \overline{\bm{\kappa}}^A$, (ii) calculate the Hessian matrix $\mathbb{H}^{\mr{XY}} (X,Y=V,O)$, (iii) calculate the multipliers by solving a linear equation, and (iv) finally we can derive analytical gradients.}
    \label{fig:analytical_grad}
\end{figure}

\subsection{Gradient for SA-OO-SSVQE}
For SA-OO-SSVQE, the Lagrangian for $A$th eigenstate is defined as
\begin{equation}
\begin{split}
	&L_A(\bm{x}, \bm{\theta}, \bm{\kappa}, \overline{\bm{\theta}}^A, \overline{\bm{\kappa}}^A)\\&= \bra{\psi_A}\hat{U}^\dag(\bm{\theta})\hat{\mc{H}}(\bm{x}, \bm{\kappa})\hat{U}(\bm{\theta})\ket{\psi_A}\\&+\sum_i\overline{\theta}^A_i\left(\frac{\partial E^{\mr{SSVQE}}}{\partial\theta_i}-0\right)\\&+\sum_{p>q}\overline{\kappa}_{pq}^A\left(\frac{\partial E^{\mr{SA}}}{\partial\kappa_{pq}}-0\right),
\end{split}
\end{equation}    
where $\theta_i$ and $\kappa_{pq}$ are elements of $\bm{\theta}$ and $\bm{\kappa}$, respectively.
We also define $\overline{\theta}_i^A$ and $\overline{\kappa}_{pq}^A$ as the Lagrange multipliers.
We impose extremal conditions for all parameters except for $\bm{x}$ on the Lagrangian, 
\begin{equation}\label{Seq: ssqve extremal}
\pdv{L_A}{\theta_j} = \pdv{L_A}{\kappa_{mn}} =
  \pdv{L_A}{\overline{\theta}_j} = \pdv{L_A}{\overline{\kappa}_{pq}} = 0.
\end{equation}
The last two equations are satisfied by taking $\bm{\theta}$ and $\bm{\kappa}$ as the optimized parameters of the SA-OO-SSVQE calculation, $\bm{\theta}^*$ and $\bm{\kappa}^*$, respectively.
The first two equations are simplified to a linear equation by defining the following vectors and matrices,
\begin{equation}\label{eq:vqe matrix}
\begin{split}
    f_j^A&=\frac{\partial E_A}{\partial\theta_j},\,\,g^A_{(mn)}=\frac{\partial E_A}{\partial\kappa_{mn}},\\\mathbb{H}^{\mr{VV}}_{ji}&= \frac{\partial^2E^{\mr{SSVQE}}}{\partial\theta_i\partial\theta_j},\,\,
    \mathbb{H}^{\mr{VO}}_{j(pq)}=\frac{\partial^2E^{\mr{SA}}}{\partial\kappa_{pq}\partial\theta_j},\\
    \mathbb{H}^{\mr{OV}}_{(mn)i}&=\frac{\partial^2E^{\mr{SSVQE}}}{\partial\theta_i\partial\kappa_{mn}},\,\,\mathbb{H}^{\mr{OO}}_{(mn)(pq)}=\frac{\partial^2E^{\mr{SA}}}{\partial\kappa_{pq}\partial\kappa_{mn}}, 
\end{split}
\end{equation}
where $ E_A =\bra{\psi_A}\hat{U}^\dag(\bm{\theta})\hat{\mc{H}}(\bm{x},\bm{\kappa})\hat{U}(\bm{\theta})\ket{\psi_A}$.
Specifically, we have 
\begin{equation}\label{matrix_eqs}
    \begin{pmatrix}
    \mathbb{H}^{\mr{VV}} & \mathbb{H}^{\mr{VO}}\\ \mathbb{H}^{\mr{OV}} & \mathbb{H}^{\mr{OO}}
    \end{pmatrix}
    \begin{pmatrix}
    \overline{\bm{\theta}}^A\\ \overline{\bm{\kappa}}^A
    \end{pmatrix}
    = -\begin{pmatrix}
    \bm{f}^A \\\bm{g}^A
    \end{pmatrix}.
\end{equation}
All the elements of $\mathbb{H}^\mr{VV,VO,OV,OO}$ and $\bm{f}^A, \bm{g}^A$ can be computed by simple measurements on quantum computers that are expected to be executable on the NISQ devices (see the following section in SI).

Once the values of the Lagrange multipliers $\overline{\bm{\theta}}^A, \overline{\bm{\kappa}}^A$ are determined by solving the linear equation~\eqref{matrix_eqs} (which we denote $\overline{\bm{\theta}}^{A*}, \overline{\bm{\kappa}}^{A*}$),
the analytical gradient of the energy $E_A$ can be computed by leveraging the extremal conditions of the Lagrangian.
That is, for solutions to the extremal condition on $L_A$, we have $L_A(\bm{x}, \bm{\theta}^*, \bm{\kappa}^*, \overline{\bm{\theta}}^{A*}, \overline{\bm{\kappa}}^{A*}) = E_A^*(\bm{x})$,
where $E_A^*(\bm{x})$ is the energy obtained by SA-OO-SSVQE.
It is straightforward to show
\begin{strip}
\begin{equation}
\begin{split}
\dv{E_A^*(\bm{x})}{x_\mu}
& = \dv{L_A(\bm{x}, \bm{\theta}^*, \bm{\kappa}^*, \overline{\bm{\theta}}^{A*}, \overline{\bm{\kappa}}^{A*})}{x_\mu} 
= \left( \pdv{x_\mu} +  \pdv{\bm{\theta}_A}{x_\mu} \cdot \bm{\nabla}_{\bm{\theta}_A} + \pdv{\bm{\kappa}}{x_\mu} \cdot \bm{\nabla}_{\bm{\kappa}}+\pdv{\overline{\bm{\theta}}^A}{x_\mu} \cdot \bm{\nabla}_{\overline{\bm{\theta}}^A}
+ \pdv{\overline{\bm{\kappa}}^A}{x_\mu} \cdot \bm{\nabla}_{\overline{\bm{\kappa}}^A} \right) L_A \\
& =\pdv{L_A(\bm{x}, \bm{\theta}^*, \bm{\kappa}^*, \overline{\bm{\theta}}^{A*}, \overline{\bm{\kappa}}^{A*})}{x_\mu},
\end{split}
\label{eq:total derivative in energy = partial derivative in Lagrangian}
\end{equation}
or more explicitly,
\begin{equation} \label{eq:inter_expression}
 \begin{aligned}
  &\dv{E_A^*(\bm{x})}{x_\mu} = \braket{\psi_A | \hat{U}^\dag(\bm{\theta}^*) \pdv{\hat{\mc{H}}(\bm{x},\bm{\kappa}^*)}{x_\mu} \hat{U}(\bm{\theta}^*) | \psi_A}  + \sum_i\sum_S \overline{\theta}_i^{A*} \omega_S^{\mr{VQE}}\pdv{\theta_i} \bra{\psi_S}\hat{U}^\dag(\bm{\theta}^*)  \pdv{\hat{\mc{H}}(\bm{x},\bm{\kappa}^*)}{x_\mu} \hat{U}(\bm{\theta}^*)\ket{\psi_S}\\
 & + \sum_{p>q}\sum_S \overline{\kappa}_{pq}^{A*} \omega_S^{\mr{SA}} \pdv{\kappa_{pq}} \bra{\psi_S} \hat{U}^\dag(\bm{\theta}^*) \pdv{\hat{\mc{H}}(\bm{x},\bm{\kappa}^*)}{x_\mu} \hat{U}(\bm{\theta}^*)\ket{\psi_S}.
 \end{aligned}
\end{equation}
\end{strip}
Again, the values on the right-hand side can be computed by quantum computers in the way described in the next section.

\subsection{Gradient for SA-OO-MCVQE}
As reviewed in the previous section, the energies for SA-OO-MCVQE are determined by the eigenvalue problem, Eq.~\eqref{Seq: mcvqe eigeneqn}.
When the Hamiltonian $\hat{\mc{H}}$ has $\bm{x}$ dependence, the matrix $h_{ST}$, the vector $v^{(A)}_S$, and the energy $E_A^*$ also have $\bm{x}$ dependence.
By differentiating Eq.~\eqref{Seq: mcvqe eigeneqn} with respect to $x_\mu$ and using $\sum_S (v^{(A)}_S)^* v^{(B)}_S = \delta_{AB}$, we find
\begin{equation}\begin{split} \label{eq: grad MCVQE}
 \dv{E_A^*(\bm{x})}{x_\mu}= \sum_{ST}\left(v_S^{(A)}\right)^*\frac{dh_{ST}}{dx_\mu}v_T^{(A)}.
 \end{split}
\end{equation}
Therefore, the gradient of the energy $E_A^*$ is obtained by calculating the derivative of the matrix elements $h_{ST}$.

To calculate $\dv{h_{ST}}{x_\mu}$, we consider the following Lagrangian,
\begin{equation}\begin{split} \label{eq: Lagrangian MCVQE}
 \begin{aligned}
 & L_{ST}(\bm{x}, \bm{\kappa}, \bm{\theta}, \overline{\bm{\theta}}^{ST}, \overline{\bm{\kappa}}^{ST})\\& = h_{ST}(\bm{x}, \bm{\kappa}, \bm{\theta}) + \sum_i\overline{\theta}^{ST}_i\left(\frac{\partial E^{\mr{MCVQE}}}{\partial\theta_i}-0\right)\\&+\sum_{p>q}\overline{\kappa}^{ST}_{pq}\left(\frac{\partial E^{\mr{SA}}}{\partial\kappa_{pq}}-0\right),
 \end{aligned}
 \end{split}
\end{equation}
where $h_{ST}(\bm{x}, \bm{\kappa}, \bm{\theta}) = \braket{\psi_S|\hat{U}^\dag(\bm{\theta}) \hat{\mc{H}}(\bm{x}, \bm{\kappa}) \hat{U}(\bm{\theta})|\psi_T}$.
We extremize the Lagrangian with respect to the parameters except for $\bm{x}$,
\begin{equation}
 \pdv{L_{ST}}{\theta_j} = \pdv{L_{ST}}{\kappa_{mn}} = \pdv{L_{ST}}{\overline{\theta}_j} = \pdv{L_{ST}}{\overline{\kappa}_{mn}} = 0.
\end{equation}
As is the case with other methods, $\pdv{L_{ST}}{\overline{\theta}_j} = \pdv{L_{ST}}{\overline{\kappa}_{mn}} = 0$ is satisfied by choosing $\bm{\theta} = \bm{\theta}^*, \bm{\kappa} = \bm{\kappa}^*$, where $\bm{\theta}^*, \bm{\kappa}^*$ are the optimal parameters as a result of the SA-OO-MCVQE calculation.
The equations $\pdv{L_{ST}}{\theta_j} = \pdv{L_{ST}}{\kappa_{mn}} = 0$ amount to a linear equation,
\begin{equation}
 \begin{pmatrix}
    \mathbb{H}^{\mr{VV}} &\mathbb{H}^{\mr{VO}} \\
    \mathbb{H}^{\mr{OV}} &\mathbb{H}^{\mr{OO}}
 \end{pmatrix}
 \begin{pmatrix}
  \overline{\bm{\theta}}^{ST}\\\overline{\bm{\kappa}}^{ST}
 \end{pmatrix}
 =-\begin{pmatrix}
 \bm{f}^{ST} \\ \bm{g}^{ST}
 \end{pmatrix},
\end{equation}
where
\begin{equation}\begin{aligned}
    f_{j}^{ST}&=\frac{\partial h_{ST}}{\partial\theta_j},\,\,g_{(mn)}^{ST}=\frac{\partial h_{ST}}{\partial\kappa_{mn}}\\
    \mathbb{H}^{\mr{VV}}_{ji}&=\frac{\partial^2E^{\mr{MCVQE}}}{\partial\theta_i\partial\theta_j},\,\,\mathbb{H}^{\mr{VO}}_{j(pq)}=\frac{\partial^2E^{\mr{SA}}}{\partial\kappa_{pq}\partial\theta_j}\\
    \mathbb{H}^{\mr{OV}}_{(mn)i}&=\frac{\partial^2E^{\mr{MCVQE}}}{\partial\theta_i\partial\kappa_{mn}},\,\,\mathbb{H}^{\mr{OO}}_{(mn)(pq)}=\frac{\partial^2E^{\mr{SA}}}{\partial\kappa_{pq}\partial\kappa_{mn}}.
\end{aligned}
\end{equation}
The analytical gradient of $h_{ST}$ is computed as
\begin{strip}
\begin{equation} \label{eq:analytical grad H_ST}
 \begin{aligned}
  &\frac{dh_{ST}}{dx_\mu}=\bra{\psi_S}\hat{U}^\dag(\bm{\theta}^*)\frac{\partial\hat{\mc{H}}(\bm{x},\bm{\kappa}^*)}{\partial x_\mu}\hat{U}(\bm{\theta}^*)\ket{\psi_T} + \frac{1}{K}\sum_i\sum_S\overline{\theta}^{ST*}_i\frac{\partial}{\partial\theta_i}\bra{\psi_S}\hat{U}^\dag(\bm{\theta}^*)\frac{\partial\hat{\mc{H}}(\bm{x},\bm{\kappa}^*)}{\partial x_\mu}\hat{U}(\bm{\theta}^*)\ket{\psi_S} \\
 &+ \sum_{p>q} \sum_S \overline{\kappa}^{ST*}_{pq} \omega_S^{\mr{SA}} \pdv{\kappa_{pq}} \braket{\psi_S| \hat{U}^\dag(\bm{\theta}^*) \frac{\partial\hat{\mc{H}}(\bm{x},\bm{\kappa}^*)}{\partial x_\mu} \hat{U}(\bm{\theta}^*) |\psi_S}.
 \end{aligned}
\end{equation}
\end{strip}
By putting the value of $\dv{h_{ST}}{x_\mu}$ in Eq.~\eqref{eq: grad MCVQE}, the energy gradient can be evaluated.

\section{Evaluation of derivatives appearing in the formulas}
In this section, we present methods to evaluate various types of parameter-derivatives appearing in formals of analytical gradients.
We reduce the evaluation of them into simple measurements of expectation values of observables on quantum computers, which is easily implementable on NISQs.

When we compute the energy derivatives according to our formulas, we have to consider the expectation value of the Hamiltonian,
\begin{equation} \label{eq: def Q}
  Q(\bm{x}, \bm{\kappa}, \bm{\theta}) = \braket{\Phi|\hat{U}^\dag(\bm{\theta}) \hat{\mc{H}}(\bm{x}, \bm{\kappa}) \hat{U}(\bm{\theta}) |\Phi},
\end{equation}
and its parameter derivatives, $\partial Q/\partial x_\mu, \partial Q/\partial \kappa_{pq}$, and $\partial Q/\partial \theta_{i}$. Here $\ket{\Phi}$ is some quantum state prepared by the quantum computer.
Although we present the way to calculate only the first-order parameter derivatives of $Q$ in this section, the cross-parameter derivatives of $Q$ like $\partial^2 Q/\partial\kappa_{pq} \partial \theta_{i}$ can also be evaluated by combing the techniques for each parameter.
We note that the evaluation of the transition amplitude like $\braket{\Psi_S|\hat{\mc{H}}|\Psi_T}$ also reduces to that of the expectation value by taking $\ket{\Psi} = (\ket{\Psi_S} \pm (i)\ket{\Psi_T})/\sqrt{2}$, as discussed in Refs.~\citenum{nakanishi2018subspace, Parrish2019PRL}.

\subsection{Derivative with respect to $x$}
A derivative of $Q$ with respect to $x_\mu$ is easy to evaluate on quantum computers.
To see this, let us consider $\bm{x}$-derivative of the original Hamiltonian (Eq.~(1)) 
\begin{equation}
 \begin{split}
   \dv{\hat{H}(\bm{x})}{x_\mu} &= \dv{E_\mr{c}(\bm{x})}{x_\mu} + \sum_{ij\sigma} \dv{h_{ij}(\bm{x})}{x_\mu} \hat{a}_{i\sigma}^\dag \hat{a}_{j\sigma}\\&+ \frac{1}{2}\sum_{ijkl\sigma\tau} \dv{g_{ijkl}(\bm{x})}{x_\mu} \hat{a}_{i\sigma}^\dag\hat{a}_{j\tau}^\dag\hat{a}_{k\tau}\hat{a}_{l\sigma}.
 \end{split}
\end{equation}
The derivatives of the coefficients,
\begin{equation*}
\dv{\hat{H}(\bm{x})}{x_\mu}, \dv{h_{ij}(\bm{x})}{x_\mu}, \dv{g_{ijkl}(\bm{x})}{x_\mu},
\end{equation*}
can be analytically computed by classical computers.
We note that those derivative does not contain contributions from the orbital response.
Projecting the Hamiltonian derivative into the active space $d\hat{H}/dx_\mu$ yields
\begin{equation}
 \begin{split}
 \pdv{\hat{\mc{H}}(\bm{x}, \bm{\kappa})}{x_\mu} 
 &= E^{\prime}_\mr{c}(\bm{x},\bm{\kappa}) + \sum_{ij\sigma}^\mr{AS} h_{ij}^\prime (\bm{x},\bm{\kappa}) \hat{a}_{i\sigma}^\dag \hat{a}_{j\sigma}\\& + \frac{1}{2}\sum_{ijkl\sigma\tau}^\mr{AS} g_{ijkl}^\prime(\bm{x},\bm{\kappa}) \hat{a}_{i\sigma}^\dag\hat{a}_{j\tau}^\dag\hat{a}_{k\tau}\hat{a}_{l\sigma},
 \end{split}
\end{equation}
where ``AS" means that the summation of the MO indices $i,j,k,l$ is taken within the active space.
The values of $E^{\prime}_\mr{c}(\bm{x},\bm{\kappa}), h_{ij}^\prime (\bm{x},\bm{\kappa})$ and $g_{ijkl}^\prime(\bm{x},\bm{\kappa})$ can be easility computed by classical computers.
The derivative of $Q$ with respect to $x_\mu$ is then written as
\begin{equation}
 \begin{split}
 &\pdv{Q(\bm{x}, \bm{\kappa}, \bm{\theta})}{x_\mu} \\&= E^{\prime}_\mr{c}(\bm{x},\bm{\kappa}) + \sum_{ij\sigma}^\mr{AS} h_{ij}^\prime (\bm{x},\bm{\kappa}) \braket{\Phi|\hat{a}_{i\sigma}^\dag \hat{a}_{j\sigma}|\Phi} \\
 &+ \frac{1}{2}\sum_{ijkl\sigma\tau}^\mr{AS} g_{ijkl}^\prime(\bm{x},\bm{\kappa}) \braket{\Phi|\hat{a}_{i\sigma}^\dag\hat{a}_{j\tau}^\dag\hat{a}_{k\tau}\hat{a}_{l\sigma}|\Phi}.
 \end{split}
\end{equation}
Therefore, evaluating the expectation values of
$\braket{\Phi|\hat{a}_{i\sigma}^\dag \hat{a}_{j\sigma}|\Phi}$ and $\braket{\Phi|\hat{a}_{i\sigma}^\dag\hat{a}_{j\tau}^\dag\hat{a}_{k\tau}\hat{a}_{l\sigma}|\Phi}$, or 1,2-RDMs, suffices to obtain the derivative of $Q$ with respect to $x_\mu$.
 
\subsection{Derivative with respect to circuit parameters}
A derivative of $Q$ with respect to a circuit parameter $\theta_i$ can be evaluated with the technique called ``parameter shift rule"~\cite{Mitarai2018, Schuld2019, izmaylov2021analytic}.
For simplicity, we explain it for the simplest case where the ansatz circuits consist of sequence of the Pauli rotation gate, i.e.,
\begin{equation}
\label{eq: form of ansatz}
 \begin{split}
 \hat{U}(\bmth) &=
 \prod_{\hat{P}_j\in \{X, Y, Z, I\}^{\otimes N}} \exp\left[ - i\frac{\theta_j}{2} \hat{P}_j \right] \\
 &\equiv \hat{U}_M(\theta_M) \cdots \hat{U}_2(\theta_2) \hat{U}_1(\theta_1)
 \end{split}
\end{equation}
where $X, Y, Z$ are Pauli matrices, $I$ is the identity operator, $N$ is the number of qubits, and $M$ is the number of the circuit parameters.
This type of ansatz is common in applications of VQE to quantum chemistry, e.g., the qubit coupled-cluster method~\cite{Ryabinkin2018}.
The parameter shift rule enables us to replace the derivative of $Q$ with respect to the circuit parameter $\theta_i$ with the expectation values of $Q$ at ``shifted" parameters.
That is,
\begin{equation}\label{eq:theta derivative}
 \begin{aligned}
 &\pdv{Q(\bm{x}, \bm{\kappa}, \bm{\theta})}{\theta_i} \\
 &= \frac{1}{2}\bra{\Phi}\hat{U}_{i;+}^\dag(\bm{\theta}) \hat{\mathcal{H}}(\bm{x}, \bm{\kappa}) \hat{U}_{i;+}(\bm{\theta})\ket{\Phi} \\
 &-\frac{1}{2}\bra{\Phi}\hat{U}_{i;-}^\dag(\bm{\theta}) \hat{\mathcal{H}}(\bm{x}, \bm{\kappa}) \hat{U}_{i;-}(\bm{\theta})\ket{\Phi},
 \end{aligned}
\end{equation}
where $\hat{U}_{i;\pm}(\bm{\theta})$ is defined as 
\begin{equation}
    \hat{U}_{i;\pm}(\bm{\theta})\equiv\left(\prod_{k>i}\hat{U}_k(\theta_k)\right)\hat{U}_i\left(\theta_i\pm\frac{\pi}{2}\right)\left(\prod_{j<i}\hat{U}_j(\theta_j)\right).
\end{equation}
It is straightforward to apply the parameter shift rule to the higher-order derivatives of $\bm{\theta}$.

\subsection{Derivatives with respect to orbital parameters}
A derivative of $Q$ with respect to the orbital parameters $\bm{\kappa}$ can also be recast into a sum of expectation values without any ancillary qubits~\cite{Yalouz_2021}.
Indeed, the evaluation can be done only by classical postprocessing after we measure the energy, or expectation value of $\hat{\mc{H}}$.

By using the definition of $\hat{U}_\mr{OO}(\bm{\kappa})$ (Eq.~(2)), 
we have the following expressions:$\qquad\qquad$
\begin{strip}
\begin{align} \label{eq:kappa derivative}
\begin{split}
&\pdv{Q(\bm{x}, \bm{\kappa}, \bm{\theta})}{\kappa_{pq}} \big|_{\bm{\kappa}=\bm{0}} 
=\bra{\Phi}\hat{U}^\dag(\bm{\theta}) \mc{P} \left[ \hat{\kappa}_{pq}, \hat{U}_\mr{OO}^\dag(\bm{\kappa})\hat{H}(\bm{x}) \hat{U}_\mr{OO}(\bm{\kappa}) \right] \mc{P} \hat{U}(\bm{\theta})\ket{\Phi} \big|_{\bm{\kappa}=\bm{0}},
\end{split} \\
\begin{split}
&\pdv{Q(\bm{x}, \bm{\kappa}, \bm{\theta})}{\kappa_{pq}}{\kappa_{mn}} \big|_{\bm{\kappa}=\bm{0}} 
= \frac{1}{2} \bra{\Phi}\hat{U}^\dag(\bm{\theta}) \mc{P} \left[\hat{\kappa}_{mn}, \left[ \hat{\kappa}_{pq}, \hat{U}_\mr{OO}^\dag(\bm{\kappa}) \hat{H}(\bm{x}) \hat{U}_\mr{OO}(\bm{\kappa}) \right] \right] \mc{P} \hat{U}(\bm{\theta})\ket{\Phi} \big|_{\bm{\kappa}=\bm{0}} \\
&+ \frac{1}{2} \bra{\Phi}\hat{U}^\dag(\bm{\theta}) \mc{P} \left[\hat{\kappa}_{pq}, \left[ \hat{\kappa}_{mn}, \hat{U}_\mr{OO}^\dag(\bm{\kappa}) \hat{H}(\bm{x}) \hat{U}_\mr{OO}(\bm{\kappa}) \right] \right] \mc{P} \hat{U}(\bm{\theta})\ket{\Phi} \big|_{\bm{\kappa}=\bm{0}},
\end{split}
\end{align}
\end{strip}
where $\hat{\kappa}_{pq}$ is defined as 
\begin{equation}
 \hat{\kappa}_{pq} = \sum_\sigma (\hat{a}^\dag_{p\sigma}\hat{a}_{q\sigma}-\hat{a}_{q\sigma}^\dag\hat{a}_{p\sigma}).
\end{equation}
Because of the form of the operator $\hat{\kappa}_{pq}$,
the number of annihilation and creation operators in the commutators $[\hat{\kappa}_{pq}, \hat{\mc{H}}]$, $[\hat{\kappa}_{mn},  [\hat{\kappa}_{pq}, \hat{\mc{H}}]]$ is the same as that of the original $\hat{\mc{H}}$.
Therefore, by classically processing the commutators and combining the result of it with the expectation values of the creation and annihilation operators, we can evaluate the derivatives of $Q$ with respect to $\bm{\kappa}$.
In the following, we see this explicitly.

Let us recall the definition of the one-electron reduced density matrix (1-RDM) and the two-electron reduced density matrix (2-RDM),
\begin{equation}
 \begin{aligned}
 \rho_{ij}^{(1)} &\equiv \sum_\sigma\bra{\Phi}\hat{U}^\dagger(\bm{\theta})\hat{a}_{i\sigma}^\dagger\hat{a}_{j\sigma}\hat{U}(\bm{\theta})\ket{\Phi}, \\
 \rho_{ijkl}^{(2)}&\equiv\sum_{\sigma\sigma'}\bra{\Phi}\hat{U}^\dagger(\bm{\theta})\hat{a}_{i\sigma}^\dagger\hat{a}_{j\sigma'}^\dagger\hat{a}_{k\sigma'}\hat{a}_{l\sigma}\hat{U}(\bm{\theta})\ket{\Phi},
  \end{aligned} \label{eq:def of RDMs} 
\end{equation}
respectively.
Those RDMs can be evaluated on quantum computers and are the foundation to determine the energy in the SA-OO calculation.  
We write the Hamiltonian in full space as [Eq.~(1) 
in the main text]
\begin{equation}
 \begin{split}
&\hat{U}_\mr{OO}^\dag(\bm{\kappa}) \hat{H}(\bm{x}) \hat{U}_\mr{OO}(\bm{\kappa}) \\&= E_\mr{c}(\bm{x}) + \sum_{ij\sigma} h_{ij} (\bm{x}) \hat{a}_{i\sigma}^\dag \hat{a}_{j\sigma}\\&+\frac{1}{2}\sum_{ijkl\sigma\tau} g_{ijkl}(\bm{x}) \hat{a}_{i\sigma}^\dag\hat{a}_{j\tau}^\dag\hat{a}_{k\tau}\hat{a}_{l\sigma}.
 \end{split}
\end{equation}
For the second term of this Hamiltonian, it holds
\begin{equation}\label{eq:[kappa,H1]}
 \begin{aligned}
 &\sum_{ij}\sum_{\sigma} \braket{\Phi|\hat{U}^\dag(\bm{\theta}) \left[\hat{\kappa}_{pq}, {h}_{ij}\hat{a}_{i\sigma}^\dagger\hat{a}_{j\sigma}\right] \hat{U}(\bm{\theta}) |\Phi}\\ &=-\sum_k {h}_{kp}\rho^{(1)}_{kq}+\sum_k {h}_{kq}\rho_{kp}^{(1)}\\&+\sum_k {h}_{qk}\rho_{pk}^{(1)}-\sum_k {h}_{pk}\rho_{qk}^{(1)}.
 \end{aligned}
\end{equation}
For the third term, we obtain
\begin{equation}\label{eq:[kappa,H2]}
 \begin{aligned}
 &\sum_{ijkl}\sum_{\sigma\sigma'}\braket{\Phi|\hat{U}^\dag(\bm{\theta}) \left[\hat{\kappa}_{pq}, {g}_{ijkl}\hat{a}_{i\sigma}^\dagger\hat{a}_{j\sigma'}^\dagger\hat{a}_{k\sigma'}\hat{a}_{l\sigma}\right]\hat{U}(\bm{\theta})|\Phi}\\
 &=2\sum_{jkl} {g}_{qjkl}\rho_{pjkl}^{(2)}-2\sum_{ijl} {g}_{ijpl}\rho_{ijql}^{(2)}\\&+2\sum_{jkl} {g}_{qjkl}\rho_{qjkl}^{(2)}-2\sum_{ijl} {g}_{ijql}\rho^{(2)}_{ijpl}.
 \end{aligned}
\end{equation}
Combining Eq.~\eqref{eq:[kappa,H1]} and Eq.~\eqref{eq:[kappa,H2]}, we obtain $\partial Q/\partial\kappa_{pq}$.
The second-order derivatives can also be evaluated in the same way.

\section{\label{sec:estimation} Estimation of the number of quantum circuits to evaluate derivative}
In this section, we investigate the scaling of the number of quantum circuits to be measured for evaluating analytic gradients apart from the number of them to determine the energy and optimized wavefunction in SA-OO-VQD.
We assume that one- and two-body reduced density matrix (1-RDM and 2-RDM~\eqref{eq:def of RDMs}, respectively) at the optimal parameters are already measured.
More precisely, we assume that the optimal parameters $\bm{\kappa}^*, \bm{\theta}_0^*, ..., \bm{\theta}_{K-1}^*$ for SA-OO-VQD are already obtained and 1,2-RDMs for all the state $\ket{\psi(\bmth_S)} = U(\bmth_S)\ket{\psi}$,
\begin{equation}
 \begin{aligned}
 \rho_{ij}^{(1),S} &\equiv \sum_\sigma \bra{\psi}\hat{U}^\dagger(\bm{\theta}_S)\hat{a}_{i\sigma}^\dagger\hat{a}_{j\sigma}\hat{U}(\bm{\theta}_S)\ket{\psi}, \\
 \rho_{ijkl}^{(2),S}&\equiv\sum_{\sigma\sigma'}\bra{\psi}\hat{U}^\dagger(\bm{\theta}_S)\hat{a}_{i\sigma}^\dagger\hat{a}_{j\sigma'}^\dagger\hat{a}_{k\sigma'}\hat{a}_{l\sigma}\hat{U}(\bm{\theta}_S)\ket{\psi},
  \end{aligned} \label{eq:def of RDMs state S} 
\end{equation}
are already evaluated on quantum computers ($S=0,...,K-1$).
We denote the number of molecular orbitals in the active space by $\tilde{N}$ and the number of qubits used in a quantum computer by $N=2\tilde{N}$.

To calculate the analytical gradient of all eigenenergies of the states $A=0,...,K-1$, we first solve the linear equation (Eq.~(6)) in the main text, 
so $\mathbb{H}^{\rm{VV,VO,OV,OO}}$ and $\bm{g}^A$ need to be measured. 
For simplicity, we assume that the ansatz $U(\bmth)$ has the form of Eq.~\eqref{eq: form of ansatz}, so the number of elements in the circuit parameters is $M$.

\begin{itemize}
    \item For the evaluation of $\mathbb{H}^{\rm{VV}}$, we essentially need $\pdv{}{\theta_{Si}}{\theta_{Tj}} \bra{\psi}U^\dag(\bmth_B)\mc{H}(\bm{x},\bmkp)U(\bmth_B)\ket{\psi}$ for $S,T,B=0,...,K-1$ and $i,j=1,...,M$, which is reduced to the evaluation of the second order derivative of 1,2-RDMs with respect to the circuit parameters,
    \begin{equation*}
     \pdv{}{\theta_{Si}}{\theta_{Sj}}\rho_{pq}^{(1),S}, \pdv{}{\theta_{Si}}{\theta_{Sj}} \rho_{pqrs}^{(2),S}
    \end{equation*}
    for $S=0,...,K-1$ and $p,q,r,s=1,...,\tilde{N}=N/2$, by using the discussion in the previous section.
    Note that $\pdv{}{\theta_{Si}}{\theta_{Tj}} \bra{\psi}U^\dag(\bmth_B)\mc{H}(\bm{x},\bmkp)U(\bmth_B)\ket{\psi} = 0$ unless $S=T=B$.
    The parameter shift rule explained in the previous section requires $4M$ different quantum circuits to evaluate the second order derivative of each RDM with fixed $p,q,r,s$.
    Because $p,q,r,s$ run from 1 to $\tilde{N}$ and 1,2-RDMs with fixed $p,q,r,s$ are composed of $\order{1}$ observables in the qubit representation, the total number of different circuits to evaluate above-mentioned derivatives of 1,2-RDMs for all $S,T,i,j,p,q,r,s,A$, that is, all elements of $\mathbb{H}^{VV}$, is $\order{K\times 4M \times (N/2)^4}$ in the leading order of $N$.
    \item $\mathbb{H}^{\rm{VO}}$ and $\mathbb{H}^{\rm{OV}}$ are similarly obtained by the first order derivative of 1,2-RDMs with respect to the circuit parameters,
    \begin{equation*}
     \pdv{}{\theta_{Si}}\rho_{pq}^{(1),S}, \pdv{}{\theta_{Si}} \rho_{pqrs}^{(2),S},
    \end{equation*}
    because the derivative with respect to $\bmkp$ does not require any new 1,2-RDMs.
    The evaluation of them can be done by the parameter shift rule again, which requires $2M$ different quantum circuit for fixed $p,q,r,s$.
    The total number of different quantum circuits to determine all elements of $\mathbb{H}^{VO}$ and $\mathbb{H}^{OV}$ is $\order{K\times 2M \times (N/2)^4}$.
    \item Evaluation of $\mathbb{H}^{\rm{OO}}$ and $\bm{g}^A$ can be performed with only 1,2-RDMs of the state $A$ and does not require any new quantum circuits when one uses the method in the previous section. 
    \end{itemize}
In total, the number of additional circuits to solve the linear equation~(6) in the main text and hence calculate the Lagrange multipliers $\bar{\bmkp}^A$ and $\bar{\bmth}_S^A$ is $\order{KMN^4}$.

The obtained values of the Lagrange multipliers are plugged into Eq.~(8) in the main text to calculate the analytical derivative of SA-OO-VQD energy $E^*_A$.
Since we already obtained 1,2-RDMs of the state $A$ as well as their first order derivative with respect to the circuit parameters in the course of calculating the Lagrange multipliers,
calculation of Eq.~(8) does not require any additional quantum circuits to be measured and sole classical post-processing is sufficient.
Therefore, the scaling of the number of additional quantum circuits to calculate the analytical derivative of SA-OO-VQD energy (Eq.~(8)) is $\order{KMN^4}$.

\section{Computational details of numerical experiment} 
In the numerical experiment, we have calculated the three lowest singlet states (S$_0$, S$_1$, and S$_2$) of TFP molecule using SA-OO-VQD (quantum approach) and its equivalent SA-CASSCF method (classical approach). 

In the quantum approach, we employ the real-valued symmetry preserving (RSP)~\cite{Gard2019, ibe2020calculating} with $D=10$ as a variational quantum circuit for the ansatz quantum circuit $U(\bmth)$(Fig.~\ref{fig:RSP}).
\begin{figure}
    \centering
    \includegraphics[width=.48\textwidth]{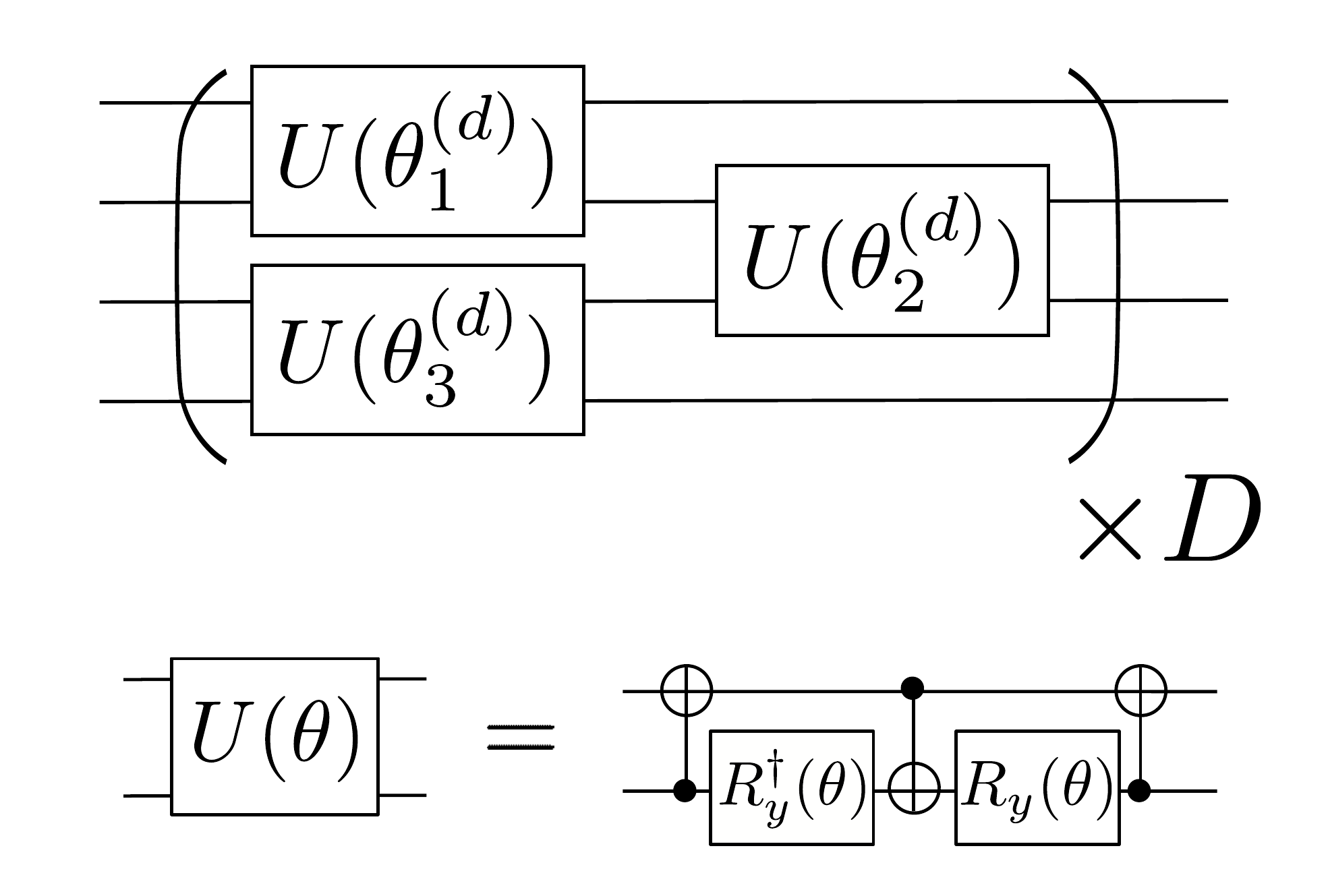}
    \caption{A diagram of real-valued symmetry preserving ansatz (RSP). $R_y(\theta)=\exp\left(-i\theta/2Y\right)$ is a single-qubit rotation gate about $y$ axis with a variational parameter $\theta$. $D$ denotes the depth of the ansatz.}
    \label{fig:RSP}
\end{figure}
The total number of the parameters is thirty.
The hyperparameter for VQD is $\beta_1 = 1, \beta_0 = 4$.
To obtain the singlet states, we add the penalty term $\beta_{\hat{S}^2}(\hat{S}^2 - 0)^2$ with $\beta_{\hat{S}^2} = 10$ to the cost function.
The optimization of the cost function of VQD [Eq.~(3)] 
is performed by BFGS method implemented in \verb|SciPy| library.
The initial values of the circuit parameters are taken as uniform random numbers drawn from $[0, 2\pi]$ when we start the S$_1$/S$_1$ CI$_{\rm MIN}$ optimization and the calculation of MEPs.
When we calculate the successive points (molecular structures) in these calculations, the optimized parameters at the previous point are used as initial parameters.
As mentioned in the main text, we assume no noise in quantum circuits and measurements, so expectation values of observable $\hat{O}$ for a state $\ket{\Phi}$ are calculated exactly as $\ev{\hat{O}}{\Phi}$.
 The simulation of quantum circuits and measurements are run by using Qulacs~\cite{qulacs_2018}.
The orbital optimization is performed with PySCF~\cite{Sun2018_pyscf}.

As for the characterization of the S$_1$/S$_0$ CI$_{\rm MIN}$, in the quantum approach,
we introduce a phase shift parameter ($\beta$) in Eq.~(9) 
that corresponds to a rotation angle of the unitary transformation of the two degenerate states,
\begin{equation}
\begin{split}
 E'_{\pm}(\rho,\varphi)&=
\rho\Big(s_x\cos(\varphi-\beta) + s_y\sin(\varphi-\beta)\\& 
\pm d_{gh} \sqrt{1+\Delta_{gh}\cos(2(\varphi-\beta))} \Bigr),
\end{split}
\end{equation}
We determine the four conical parameters ($s_x, s_y, d_{gh}, \Delta_{gh}$) and $\beta$ by fitting the above equation to the S$_1$/S$_0$ energies calculated at the points along a circle with a radius ($\rho$) of 0.01 amu${}^{0.5} \cdot$bohr centered at the optimized CI$_{\rm MIN}$ in the BP, which is spanned by two mutually orthogonalized vectors (GD and updated approximate DC vectors).
The obtained $\beta$ was $-4.7$ degrees. 
The two BP vectors ($\bm{g}$ and $\bm{h}$) were obtained by rotating the GD and the updated approximate DC vectors 
at the CI$_{\rm MIN}$ by $\beta$.
In the classical approach, the BP vectors and the conical parameters were obtained using the Yarkony's approach~\cite{Conical_YarkonyJPCA2001} from the GD and DC vectors at the optimized CI$_{\rm MIN}$.

As for the MEP search in the quantum approach, the MEP step length is set 0.05 $\AA$ and kept while the next MEP step energy is lower than the current one.
When the next step energy is higher than the current one, it was reduced by multiplying 0.8 until the next step energy is lower than the current one, where maximum cycle number is set at 20.
In the MEP calculation from CI$_{\rm MIN}$ on S$_0$, to follow the branching pathways into {\it cis} isomer and {\it trans} one, we used two slightly different starting points in the vicinity of CI$_{\rm MIN}$ on the BP, which are points displaced by 0.02 amu${}^{0.5}\cdot$bohr in the minus and plus directions of the GD vector from the optimized CI$_{\rm MIN}$ point.
In the classical approach, the MEP calculations were performed by the same way in the quantum approach, except that two starting points of S$_0$ MEP from CI$_{\rm MIN}$ were determined by displacements in the plus and minus directions of the $\bm{g}$ vector.

\end{suppinfo}

\bibliography{bibliography}


\end{document}